\documentclass{article}
\usepackage{amsmath,amssymb}
\usepackage[dvips]{graphicx}

\usepackage{latexsym}

\begin{document}

\pagestyle{plain} 
\setcounter{page}{1}
\setlength{\textheight}{700pt}
\setlength{\topmargin}{-40pt}
\setlength{\headheight}{0pt}
\setlength{\marginparwidth}{-10pt}
\setlength{\textwidth}{20cm}

\title{Braess like Paradox on Ladder  Network \\
 --Cases with two way bypasses-- }
\author{Norihito Toyota   and Fumiho Ogura \and Hokkaido Information University, Ebetsu, Nisinopporo 59-2, Japan \and email :toyota@do-johodai.ac.jp }
\date{}
\maketitle

\begin{abstract}
 Braess  \cite{1} has been studied about a traffic flow on a diamond type network   
and found that introducing  new edges to the networks always does not achieve the efficiency. 
Some researchers studied the Braess' paradox in similar type networks by introducing various types of cost functions. 
But whether such paradox occurs or not is not scarcely studied in complex networks except for Dorogovtsev-Mendes network\cite{2}. 
In this article, we study the paradox on Ladder type networks, as the first step to the research about Braess' paradox on Watts and Strogatz type small world network\cite{Watt1}\cite{Watt2}. 
For the purpose, we construct $4 \times 3$ models as extensions of the original Braess' models.
We analyze theoretically and numerically studied the models on Ladder networks.   
Last we give a phase diagram for (a) model, where the cost functions of bypasses are constant =0 or flow, base on two parameters $r$ and $p$ by numerical simulations. 
Simulation experiments also show some conditions that paradox can not occur.  
These facts give some sugestions for designing effective transportation networks.

 \end{abstract}
\begin{flushleft}
\textbf{keywords:}
 Braess' paradox, Small world network, Ladder network
\end{flushleft}

\section{ Introduction }
 
 When one transmits some information based on a self efficiency on some network, 
introducing  new edges to a network always does not achieve the efficiency. 
This feature is necessarily not restricted to information transmittance and 
is relevant to the flows of some physical objects as a traffic flow. 
This phenomenon is generally known as Braess' paradox  \cite{1} which 
has been investigated about a traffic flow on a diamond type network with one diagonal line (see Fig.1). 
This is due to the fact Nash flow is necessarily not the optimal flow.    

In \cite{pas}, cases where the cost on the network is symmetric have been investigated. 
Their result  shows that Braess' paradox can arise in such limited cases.  
In the cases with more general cost functions where a cost function on every edge is a different linear function each other,   
the conditions that Braess' paradox occurs have been closely investigated in Braess network configuration as Fig.1\cite{Zrer}. 
Moreover Valiant and Royghtgarden \cite{Vali}  have proved that Braess' paradox is likely to occur in  a natural random network. 
An instructive review and many references are given in \cite{Bloy} 

A study that can be interpreted as a phenomenon like Braess' paradox on a sort of small world network has been done \cite{2}. 
They analytically investigated when the average shortest path is optimal on Dorogovtsev-Mendes network\cite{2}. 
Moreover the authors in \cite{3,4} analytically and numerically studied the situation where some cost is required  
when one goes via the center of the network. 
They pointed out that increasing the bypass via the center does not reduce the average cost. 

The researches so far are with the proviso, however, that  information or some objects on a network returns to the start node. 
I have analytically and numerically studied the cases that a start point(node) of information/objects is different from a goal node 
on Dorogovtsev-Mendes network as the original Braess' paradox based on the research\cite{ToyotaB1}.
Through the research, I showed that though any Braess' like paradox does not occur when the network size becomes infinite,  
the paradoxical phenomenon appears at finite size of the network. 

In this article, we remove the central node from  Dorogovtsev-Mendes network such as the original small-world network\cite{Watt1}\cite{Watt2}. 
Moreover we give some short cuts on the network in special way,  where short cuts do not intersect each other, for simplicity. 
The resultant network topologically becomes a ladder type network with $N-1$ bypasses (short cuts).
We consider the Braess' paradox on the "Ladder network". 
Then while the network has not no longer properties of  small world networks, the studies of this network would give a  foundation 
to researches of complex networks. 
The basic setting in Ladder model is given in Fig.2. 
We study  models with four types of  cost functions on the  circumferential edges. 
For bypasses, we give three types of cost functions and so $4 \times 3$ models on the ladder network are discussed in all. 
We explore the models from both theoretical point of view and numerical point of view. 
We last give a phase diagram that shows which conditions and parameter area Braess' paradox occurs. 

\begin{figure}[ht]
\centering
\includegraphics[scale=0.2]{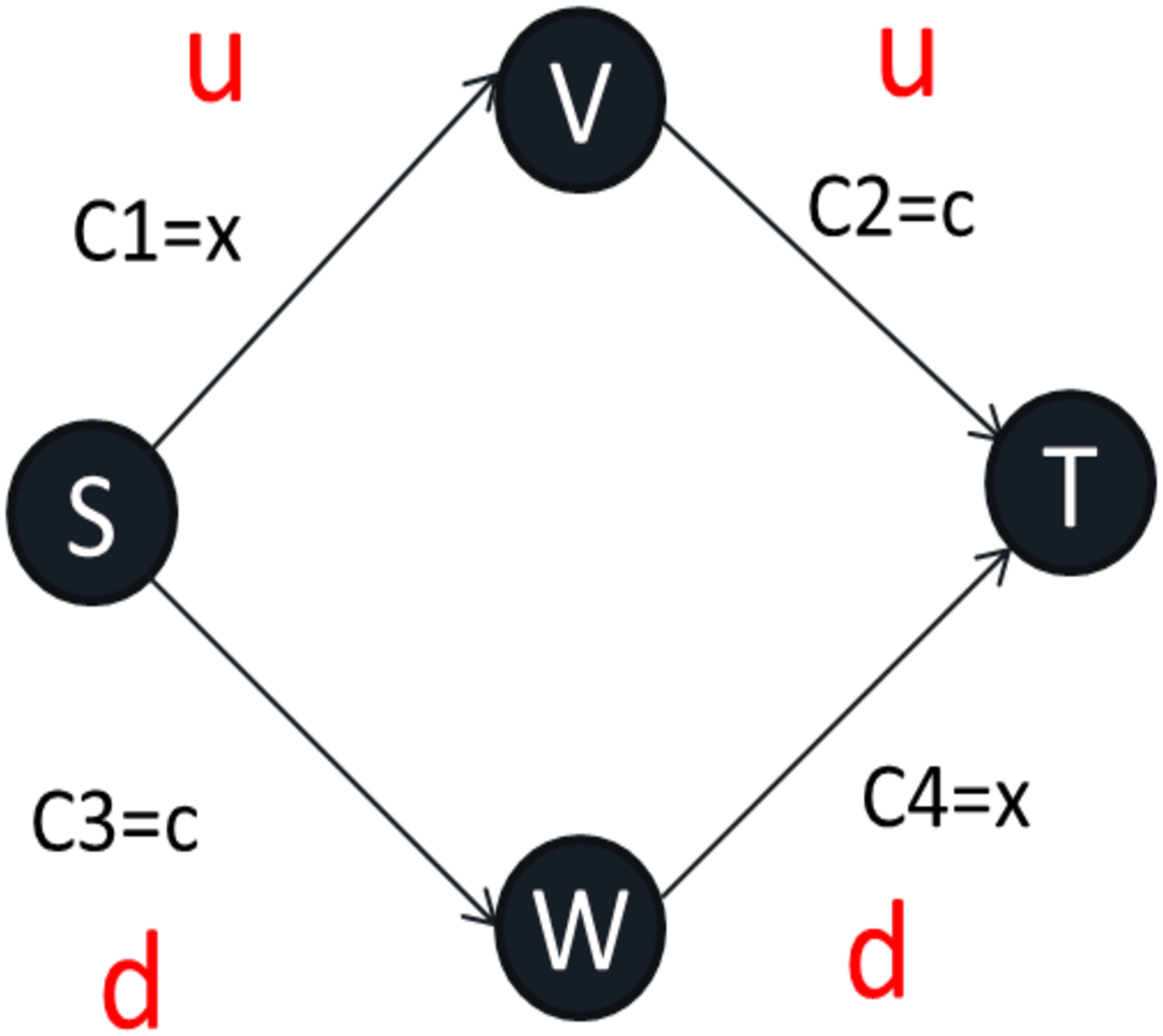}\hspace{15mm} 
\includegraphics[scale=0.2]{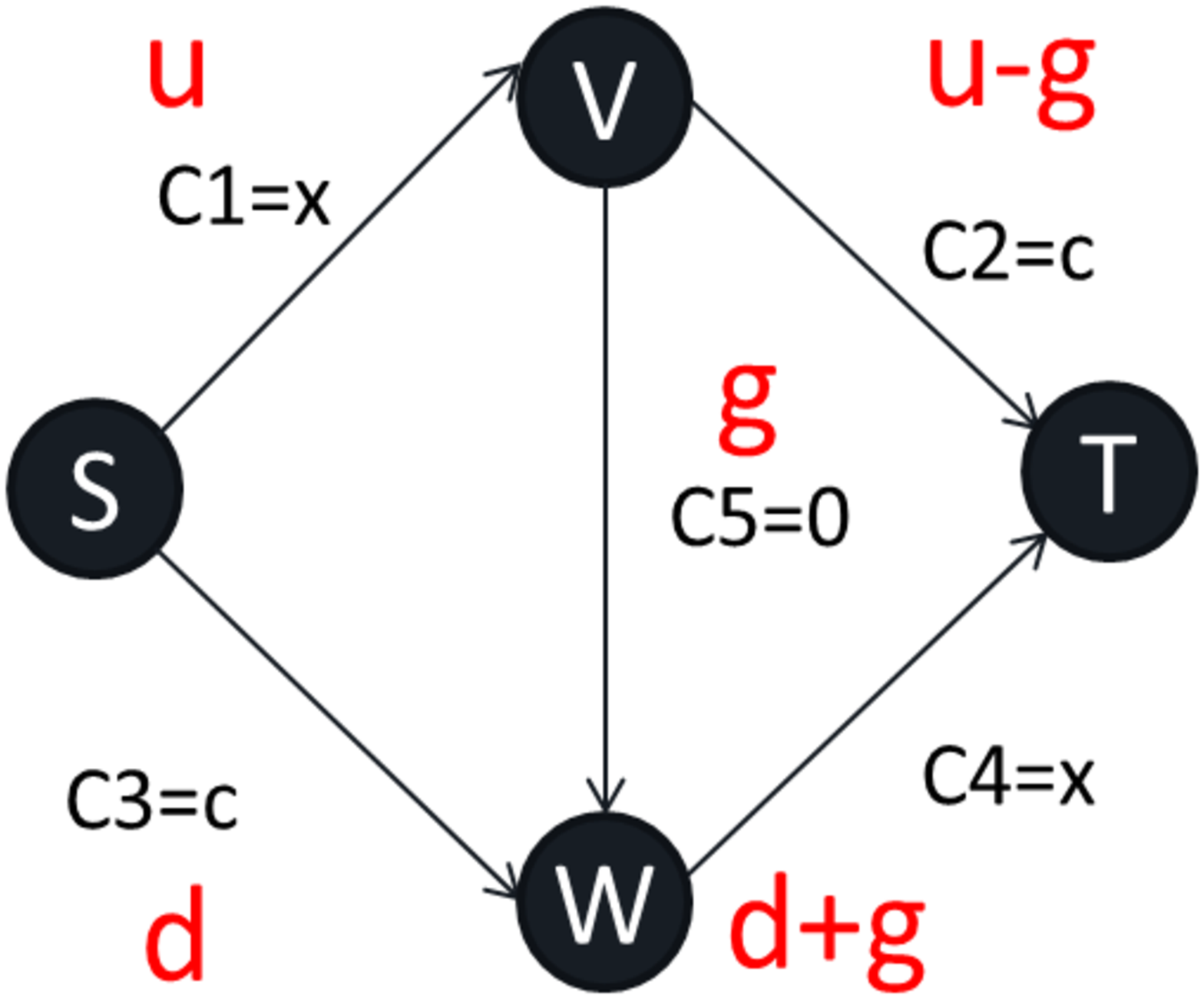}
\caption{Braess' Network}
\end{figure}

%

%

%

\section{ Models extended from the original Braess' network }
\subsection{Example of Braess' paradox}
$c_i$ is cost needed when one goes on each edge "i" in Fig.1 and $c_i=x$ means the cost is just equal to the fllow on the edge. 
The flow of an upper path is $u$ and the one of a lower path is $d=X-u$ where $X$ is the total flow that flows into the network. 
Then the Nash flow is found by solving the equation, 
which means that  both costs needed for going along the upper path and the lower path from the node S to the node T are the same. 
For the left figure of Fig.1, we obtain
\begin{equation}
u+c=c+d.
\end{equation}
So $u=d=\frac{X}{2}$ and the cost is $c+ \frac{X}{2}$. 
 For the right figure of Fig.1, we obtain the following two equations;
 \begin{equation}
u+c=c+d+g=u+0+d+g,
\end{equation}
where the rightest is the cost for the path S$\rightarrow$V$\rightarrow$W$\rightarrow$T. 
So we find that $u=c,\;d=X-c,\;g=2c-X$. 
These solutions mean that the flow of edges with the cost $c$ is $X-c$ and 
the flow of edges with the cost $=flow=x$ is $c$.   
The cost from S to T is $2c$, which is larger than the  $c+ \frac{X}{2}$ in the case with no bypass when $c > \frac{X}{2}$. 
This is an example of Braess' paradox. 
From now on, we set $c \leq X$ and $X=1$ so that $u$ and $d$ mean the ratio of the flows in this article.

When bypass is  reverse direction, we obtain $u=c$, $d=1-c$, $g=1-2c>0$ and then the cost $=2c$ by the similar argument. 
Thus there is not the paradox, because $c\leq \frac{1}{2}$ due to $g\geq 0$ and so $\frac{1}{2}+c \geq 2c$. 
When the bypass is two way traffic, there are four paths from S to T that lead to three simultaneous equations. 
Solving these equations, we obatin $u=1-d=c$, and a net traffic from V to W is $2c-1$. 
Then the cost from S to T is $2c$. 
So the paradox can occur for $\frac{1}{2}<c$.  

Next we consider $c_5=flow\; x$ in Fig.1. 
Solving similar simultaneous equations, we obatin $u=1-d=c$, and $g=2c-1$. 
So the paradox can occur for $\frac{1}{2}<c$.  
When bypass is  reverse direction, we obtain $u=\frac{1+c}{3}$, $d=\frac{2-c}{3}$, $g=\frac{1-2c}{3}$ and then the cost $=\frac{4c+1}{3}$ by the similar argument.  
The difference of the cost with a bypass and without a bypass is given by $\frac{2c-1}{6}$. 
This says that such paradox occurs for $c>\frac{1}{2}$.  
When the bypass is two way traffic, there are four paths that lead to three simultaneous equations. 
Solving these equations, we obatin $u=d=c=\frac{1}{2}$, and so there are no traffic in both ways. 
This means there is not the paradox. 
Notice that the number of simultaneous equations in this case is one more than the ones in two way traffic.  
So $c$ is determined by the equations. 
This is a crucial deference between both cases. 
All patterns with $c=1$ are given in Appendix B. 


\subsection{Models}
So far the diamond type network as shown in Fig.1 has been mainly investigated. 
It is known that complex networks are ubiquitous networks in the real world. 
It may be important to investigate Braess' paradox in complex networks. 
Some researches have been made in small world networks. 
 Dorogovtsev-Mendes network\cite{2} is a small world like network introduced by Watts and Strogatz\cite{Watt1,Watt2} 
but with one center node. 
Some edges are drawn from circumferential nodes to the center with probability $p$.    
The edges on the circumferential nodes are directed but the edges drawn from circumferential nodes to the center 
are not so.  
We have considered the total cost needed to go from a starting node to a terminal node to discuss Braess like paradox\cite{ToyotaB1},\cite{ToyotaB2}. 

\begin{figure}[ht]
\centering
\includegraphics[scale=0.4]{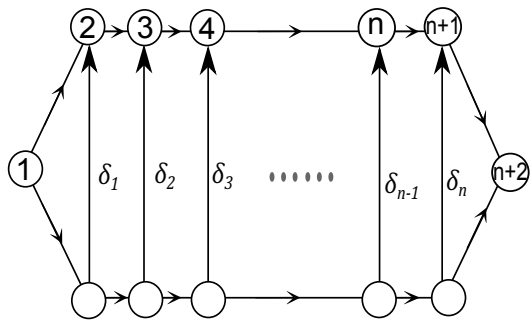}
\caption{Basic figure based on Ladder networks}
\end{figure}
In this article,   we consider Dorogovtsev-Mendes network without the center node to imitate the Watts and Strogatz type network. 
Moreover we remove the crossings in the network for simplicity.  
While the resultant network does not become so-called small world network but becomes a Ladder-type network topologically shown in Fig.2, 
it is easier to theoretically analyze it.  
This study will form the basis of the researches on Braess' paradox in the small world networks. 

 We introduce some preliminary models to uncover essential properties on the Braess like paradox in the network. 
 The networks of this type can be considered as some extensions of Braess' orginal network, a dyamond type network. 
4$\times$ 3 models discussed in this article are as follows. 
We first introduce 4 types of models for  the costs of circumferential edges\\

A. Symmetric Models

(i) P-symmetric model: 
 Both upper edges and lower edges have the constant cost $c$ with probability $1-p$ and the cost $=$ flow $x$ with the probability $p$, respectively.
 
 (ii)T-symmetric model: 
 Both upper edges and lower edges have the constant cost $c$ with probability $1-p$ and the cost $=$ flow $x$ with the probability $p$, respectively, 
 but the costs of the first upper edge and the first lower edge are fixed to the flow $=x$ on the edge and $c$. 
  The costs of the last lower edge and the last upper edge are also fixed to the flow $=x$ on the edge and $c$. \\
  
  These models are definitely given in Fig.3 and Fig.4 for (a) model, which will be expalained immediately below. \\
 
\begin{figure}[ht]
\centering
\includegraphics[scale=0.4]{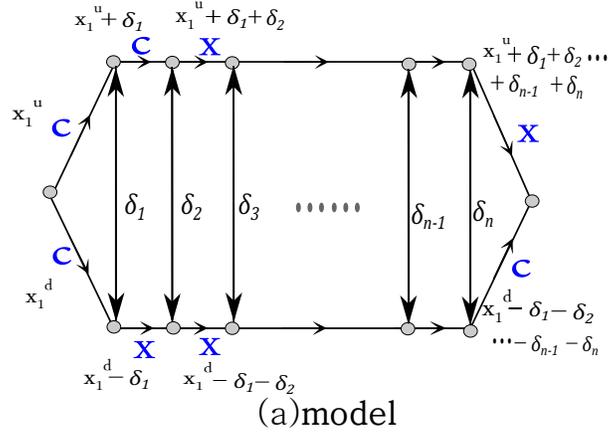}
\caption{P-symmetric model}
\end{figure}
\begin{figure}[ht]
\centering
\includegraphics[scale=0.4]{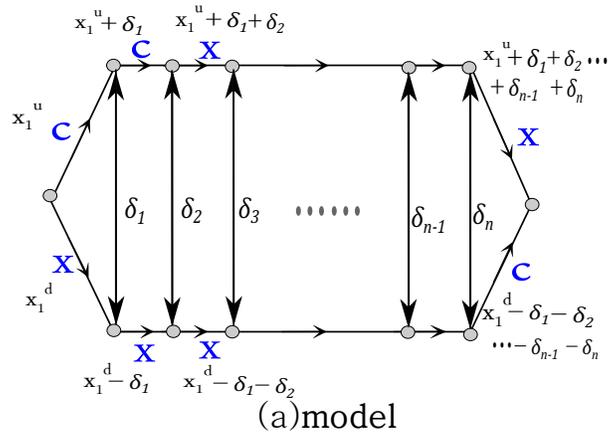}
\caption{T-symmetric model}
\end{figure}
  
B. Dual Models
\begin{figure}[ht]
\centering
\includegraphics[scale=0.4]{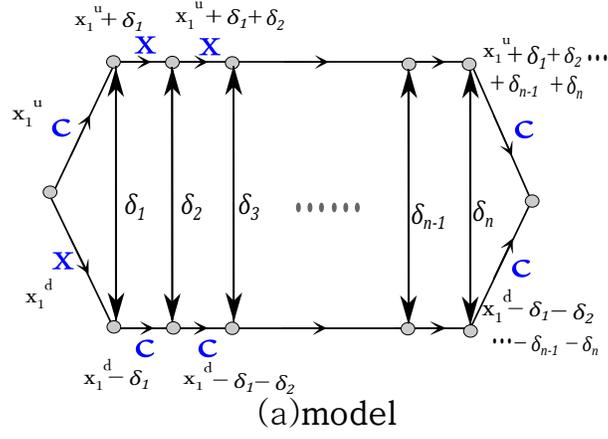}
\caption{P-dual model}
\end{figure}
\begin{figure}[ht]
\centering
\includegraphics[scale=0.4]{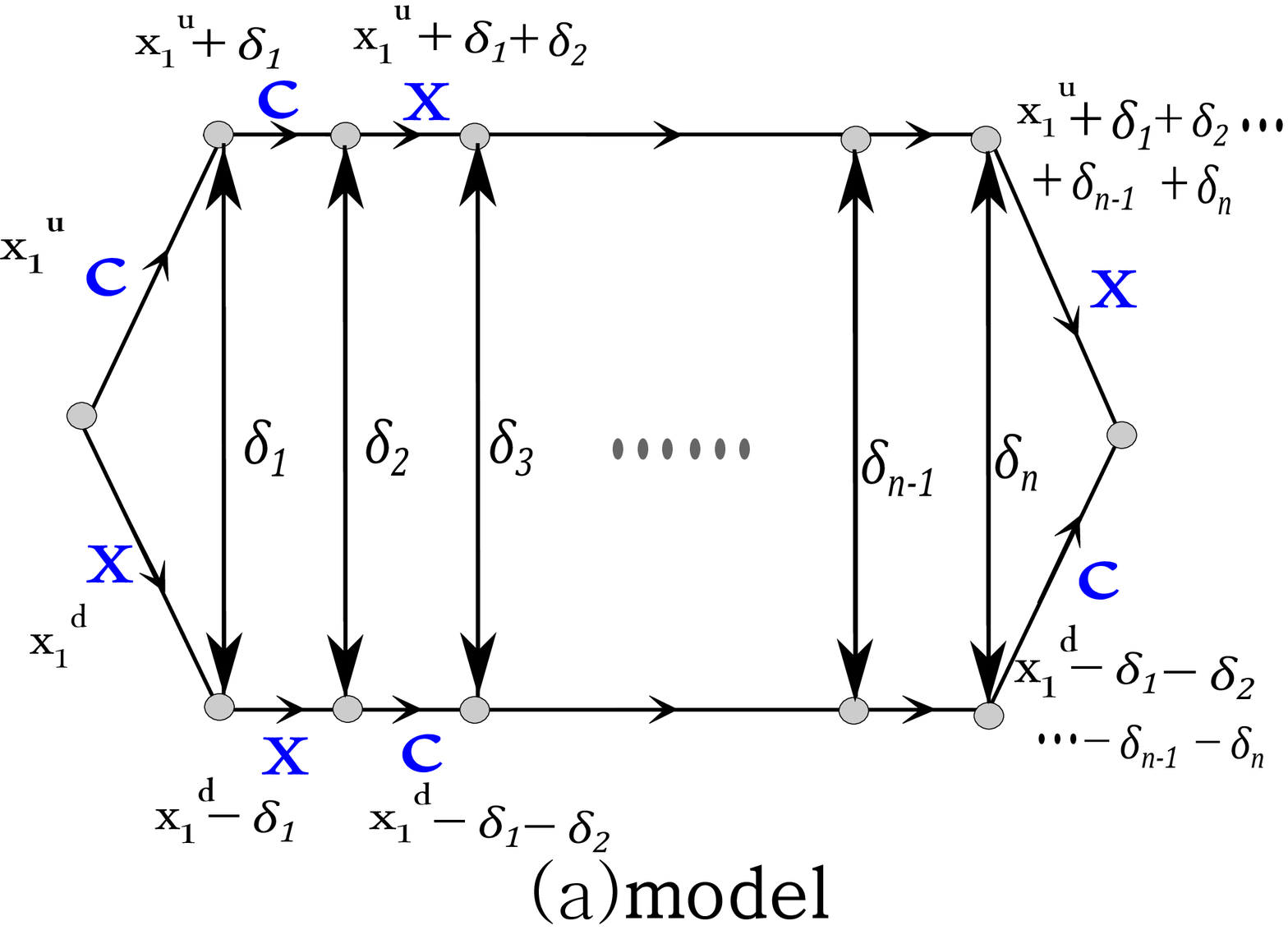}
\caption{T-dual model}
\end{figure}

  (i) S-dual model:
     The upper edges  have the constant cost $c$ with probability $1-p$ and the cost $=$ flow $x$ with the probability $p$, respectively, 
but the lower edges  have the constant cost $c$ with probability $p$ and the cost  $=$ flow with $x$ the probability $1-p$, respectively.

 (ii)T-dual model:
 The upper edges  have the constant cost $c$ with probability $1-p$ and the cost $=$ flow $x$ with the probability $p$, respectively, 
 but the lower edges have the constant cost $c$ with probability $p$ and the cost $=$ flow $x$ with the probability $1-p$, respectively.
 The costs of the first upper edge and the first lower edge are fixed to the flow $=x$ on the edge and $c$. 
  The costs of the last lower edge and the last upper edge are also fixed to the flow $=x$ on the edge and $c$. \\

 These models are definitely given in Fig.5 and Fig.6 for (a) model. 
The model (ii)  cannot be simply considered as a special case of the model (i) as discussed in the section 3.  
Especially when the cost on an upper edge is $c\; (or\;x)$ and the cost of the corresponding lower edge is the flow $x\;(or\;c)$ on the edge, 
the model is reffered to as "strong dual model". \\

For the costs of bypasses, we introduce 3-types of models.

(a)The costs of bypasses are the flow $x$ with the probability $r$ and 0 with the probablity $1-r$.

(b)The costs of bypasses are the flow $x$ with the probability $r$ and 1 with the probablity $1-r$.

(c)The costs of bypasses are the flow $x$ with the probability $r$ and c, which is the same value as the constant one of circumferential edges, 
with the probablity $1-r$.

\subsection{Notations}
In this section, we give some notations used in this article.  
\begin{itemize}
\item $N$ is the half number of nodes in the Ladder network.
\item $p$ and $r$ are the possibility with cost$=x$ where $x$ shows a flow on upper edges and bypass edges, respectively. 
\item $c$ and $c^\prime$ are  constant costs on upper edges and bypass edges independent of the flow $x$, respectively. 
\item $k=Np$ is the average number of upper edges with cost $=x$. 
\item $m=N - \ell$  where $\ell$ is a label of short cut in the ladder network.  
\item $\delta_{(\ell)} \equiv  \delta_{(\ell)} ^{(+)} -  \delta_{(\ell)} ^{(-)} $\\
 =(Flow from a lower node to an upper node)-(Flow from an upper node to a lower node)
\item  $\delta_{1,2,\cdots,k}=\displaystyle  \sum_{i=1}^{k} \delta_i$,  $\vec{\delta}= ( \delta_{(1)}, \delta_{(2)}, \cdots, \delta_{(N-1)} )$ and  $\Delta_{\ell}   \equiv  \delta_{(\ell)} ^{(+)} + \delta_{(\ell)} ^{(-)} $,  
 \end{itemize}
$x^{(u)}_i$ and $x^{(d)}_i$ are  flows on $i$-th upper edge and lower one in Fig.1$\sim$ Fig.6;
\begin{eqnarray}
x^{(u)}_i= x^{(u)}_1 +\sum_{j=1}^i \delta_{(j)}, \\
x^{(d)}_i= x^{(d)}_1 -\sum_{j=1}^i \delta_{(j)}. 
\end{eqnarray}
  $X=  x^{(u)}_1 + x^{(d)}_1$ is the total flow on the network system. 
  Two condition must be satisfied in this system;
  \begin{eqnarray}
0 \leq x^{(u)}_1 + \sum_{j=1}^{N-1}  \delta_{(j)}  \leq X=1,\\
 1=X\geq x^{(d)}_1 -\sum_{j=1}^{N-1} \delta_{(j)} \geq 0.  
\end{eqnarray}
  
We get following expressions as an average cost for the upper path in each parameter area;
\newcounter{mycount}
\begin{list}{}{\setlength{\leftmargin}{3em}
\setlength{\labelwidth}{0pt}
\setlength{\itemindent}{\labelsep}
\usecounter{mycount}
\renewcommand{\makelabel}{\arabic{mycount}.}
}

\item $m\geq k$ and $N-m \geq k$, So $N-k \geq \ell \geq k$
 \begin{equation}
 S^{(u)}_{1(\ell)} (N,p,\vec{\delta}) =  \frac{\displaystyle  \sum_{i=0}^k i \delta_{(\ell)} \cdot _{m}C_i \cdot  _{\ell}C_{k-i} }{ _N C_k}
\end{equation}
\item $m\geq k$ and $N-m\leq k$  so $ \ell\leq k$ and  $ \ell\leq N- k$
 \begin{equation}
 S^{(u)}_{2(\ell)} (N,p,\vec{\delta}) = \frac{\displaystyle \sum_{i=k-\ell}^k i \delta_{(\ell)} \cdot _{m}C_i \cdot  _{\ell}C_{k-i} }{ _N C_k}
\end{equation}

\item $m \leq k$ and $N-m \geq k$ so $\ell\geq k$ and $\ell \geq N-k$ 
 \begin{equation}
 S^{(u)}_{3(\ell)} (N,p,\vec{\delta}) = \frac{\displaystyle \sum_{i=0}^m i \delta_{(\ell)} \cdot _{m}C_i \cdot  _{\ell}C_{k-i} }{ _N C_k}
\end{equation}

\item $m\leq k$ and $N-\leq N-m$ so $N-k \leq \ell\leq k $ 
 \begin{equation}
 S^{(u)}_{4(\ell)} (N,p,\vec{\delta}) = \frac{\displaystyle  \sum_{i=k-\ell}^m i \delta_{(\ell)} \cdot _{m}C_i \cdot  _{\ell}C_{k-i} }{ _N C_k}
\end{equation}
 \end{list}
 
We get following expressions as an average cost for the lower path in each parameter area
 with  $k^\prime=N-k$
;
 \newcounter{mycountd}
\begin{list}{}{\setlength{\leftmargin}{3em}
\setlength{\labelwidth}{0pt}
\setlength{\itemindent}{\labelsep}
\usecounter{mycountd}
\renewcommand{\makelabel}{\arabic{mycountd}.}
}

\item $m\geq k^\prime$ and $N-m\geq k^\prime$ ( $N-k^\prime=\ell\geq k^\prime$)
 \begin{equation}
 S^{(d)}_{1(\ell)} (N,p,\vec{\delta}) = \frac{\displaystyle \sum_{i=0}^{k^\prime} i \delta_{(\ell)} \cdot _{m}C_i \cdot  _{\ell}C_{k^\prime-i} }{ _N C_{k^\prime}}
\end{equation}

\item $m\geq k^\prime$ and  $N-m\leq k^\prime$ ($N-k^\prime=\ell\leq k^\prime$)
 \begin{equation}
 S^{(d)}_{2(\ell)} (N,p,\vec{\delta}) = \frac{\displaystyle \sum_{i=k^\prime-\ell}^{k^\prime} i \delta_{(\ell)} \cdot _{m}C_i \cdot  _{\ell}C_{k^\prime-i} }{ _N C_{k^\prime}}
\end{equation}

\item $m\leq k^\prime$ and  $N-m\geq k^\prime$ ( $N-k^\prime=\ell\geq k^\prime$)
 \begin{equation}
 S^{(d)}_{3(\ell)} (N,p,\vec{\delta}) = \frac{\displaystyle \sum_{i=0}^m i \delta_{(\ell)} \cdot _{m}C_i \cdot  _{\ell}C_{k^\prime-i} }{ _N C_{k^\prime}}
\end{equation}

\item $m\leq k^\prime$ and  $N-m\leq k^\prime$ ($N-k^\prime=\ell\leq k^\prime$)
 \begin{equation}
 S^{(d)}_{4(\ell)} (N,p,\vec{\delta}) = \frac{\displaystyle \sum_{i=k^\prime-\ell}^m i \delta_{(\ell)} \cdot _{m}C_i \cdot  _{\ell}C_{k^\prime-i} }{ _N C_{k^\prime}}
\end{equation}
 \end{list}

Using these, we define the following quantities;
\begin{eqnarray}
S_j^{(u)}(N,p,\vec{\delta}) =\left\{
\begin{array}{ll}
\displaystyle  \sum_{\ell=k}^{N-k} S^{(u)}_{j(\ell)}(N,p,\vec{\delta})  &  for\;  j=1,\; (k\leq N/2) \\
\displaystyle  \sum_{\ell=1}^{\mbox{min} (k,N-k)} S^{(u)}_{j(\ell)}(N,p,\vec{\delta})  &  for\;  j=2, \\
\displaystyle   \sum_{\ell=\mbox{max}(k,N-k)}^{N-1} S^{(u)}_{j(\ell)}(N,p,\vec{\delta})  &  for\;  j=3, \\
\displaystyle   \sum_{\ell=N-k}^{k} S^{(u)}_{j(\ell)}(N,p,\vec{\delta})  &  for\;  j=4, \;(k\geq N/2), 
  \end{array}
\right. 
\end{eqnarray}
\begin{eqnarray}
S_j^{(d)}(N,p,\vec{\delta}) =\left\{
\begin{array}{ll}
\displaystyle  \sum_{\ell=k^\prime }^{N-k^\prime} S^{(d)}_{j(\ell)}(N,p,\vec{\delta})  &  for\;  j=1,\; (k^\prime \leq N/2) \\
\displaystyle  \sum_{\ell=1}^{\mbox{min} (k^\prime,N-k^\prime)} S^{(d)}_{j(\ell)}(N,p,\vec{\delta})  &  for\;  j=2, \\
\displaystyle   \sum_{\ell=\mbox{max}(k^\prime,N-k^\prime)}^{N-1} S^{(d)}_{j(\ell)}(N,p,\vec{\delta})  &  for\;  j=3, \\
\displaystyle   \sum_{\ell=N-k^\prime}^{k^\prime} S^{(d)}_{j(\ell)}(N,p,\vec{\delta})  &  for\;  j=4, \;(k^\prime\geq N/2) 
  \end{array}
\right. 
\end{eqnarray}
 The cost of the upper path and the cost of the lower path are represented as $\overrightarrow{C}=C(u,u,\cdots,u)$ and 
 $\underrightarrow{C}=C(d,d,\cdots,d)$, respectively.
 
 \section{Theoretical Studies}   
 In this section, we theoretically study some models given in the previous section.  
 Mainly the average cost taken in going from the start node and to the terminal node when no cost is needed in passing through  bypasses (model(a)).   
 
  \subsection{Model A(i)  (P-symmetric model)} 
    The cost needed from the start node to the terminal node  in Model A(i)  is obtained  by 
  \begin{eqnarray}
\overrightarrow{C}=N(1-p)c +Npx^{(u)}_1 + \sum_{j=1}^4 S_j^{(u)}(N,p,\vec{\delta}) \equiv  N(1-p)c +Npx^{(u)}_1 +S^{(u)}(N,p,\vec{\delta}), \label{eq:17}\\
\underrightarrow{C}=N(1-p)c +Npx^{(d)}_1 -\sum_{j=1}^4 S_j^{(d)}(N,p,\vec{\delta}) \equiv N(1-p)c +Npx^{(d)}_1 - S^{(d)}(N,p,\vec{\delta}),   \label{eq:18}
\end{eqnarray}
where the sum of $j$ is chosen and taken correctly correspondimg $m$, $k$ or $k^\prime$ and $N-m$. 
Moreover the ratio of $\sum_{j=1}^4 S_j^{(u)}(N,p,\vec{\delta})\equiv S^{(u)}$ to $\sum_{j=1}^4 S_j^{(d)}(N,p,\vec{\delta}) \equiv S^{(d)}$  is proven to be 
\begin{equation}
\alpha \equiv \frac{S^{(u)}}{S^{(d)}}= \frac{k}{N-k}=\frac{p}{1-p} \label{eq:19}
\end{equation}
in appendix A.

 In the case without any bypasses, we can find Nash flow by solving 
  $ \overrightarrow{C} =\underrightarrow{C} $ where $ S_j^{(u)}(N,p,\vec{\delta}) = S_j^{(d)}(N,p,\vec{\delta})=0 $;
  \begin{equation}
  \overrightarrow{C} -\underrightarrow{C}=Np( x^{(u)}_1- x^{(d)}_1)=0.\label{eq:20}
 \end{equation}
  So we find that 
   \begin{equation}
 x^{(u)}_1= x^{(d)}_1=X/2\label{eq:21}
 \end{equation}
is Nash fllow. 
Then the cost is 
\begin{equation}
 \overrightarrow{C} =\underrightarrow{C}=N\bigl( (1-p)c +\frac{pX}{2}\bigr).\label{eq:22}
\end{equation}
 
In the case with bypasses, we can find a necessary condition for Nash flow by 
  \begin{equation}
\overrightarrow{C} -\underrightarrow{C}=0 \Rightarrow Np( x^{(d)}_1- x^{(u)}_1)= S^{(u)}(N,p,\vec{\delta})+ S^{(d)}(N,p,\vec{\delta}).\label{eq:23}
 \end{equation}
 
 Evaluating the difference between the cost with bypasses and without bypasses, written by $\Delta C=C(bypasses) - C(no \;bypasses)$, 
we obtain a condition for Braess' paradox;
\begin{equation} 
\Delta C=\frac{Np}{2}(x_1^{(u)}-x_1^{(d)})+ S^{(u)}(N,p,\vec{\delta}) =\frac{Np}{2}(x_1^{(d)}-x_1^{(u)})- S^{(d)}(N,p,\vec{\delta})> 0.\label{eq:24}
\end{equation}
 So we speculate that the paradox does not occurs, because when $p=0$, then we find $S^{(u)}(N,p,\vec{\delta})= 0$ from (\ref{eq:19}).  
 When $p=1$, we find $S^{(d)}(N,p,\vec{\delta})=0$ from (\ref{eq:19})  and thus $\Delta C=\frac{Np}{2}(x_1^{(d)}-x_1^{(u)})$. 
The details in this case should be given by a numerical studies by a computer which will be given in the next section.   
Other cases, for example $0<p<1$, we have also to make simulations to find whether the paradox ocuurs or not.

If  $x^{(u)}_1= x^{(d)}_1$, then  we find $\Delta C=   S^{(u)}(N,p,\vec{\delta})$ from (24) and $S^{(u)}(N,p,\vec{\delta})+ S^{(d)}(N,p,\vec{\delta})=0$ from (23). 
The condition that the paradox occurs is  $ S^{(u)}(N,p,\vec{\delta})>0$, but then $ S^{(d)}(N,p,\vec{\delta})<0$ from these equations. 
This is impossible from eq.(19) that means $S^{(u)}(N,p,\vec{\delta}) $ has the same sign as $ S^{(d)}(N,p,\vec{\delta})$  for $p\geq 0$. 
Thus we speculate that paradox does not occur in this case.  

If $p=1/2$, then $S^{(u)}(N,p,\vec{\delta})= S^{(d)}(N,p,\vec{\delta}) $ from eq.(19). 
So we obtain $ N( x^{(d)}_1- x^{(u)}_1)= 4S^{(u)}(N,p,\vec{\delta}) $ from (23) for Nash flow. 
Then $ \Delta C=\frac{Np}{2}(x_1^{(u)}-x_1^{(d)})- S^{(d)}(N,p,\vec{\delta})= 0$. 
As result, it is conjectured that the paradox can not occur under these arguments.



  \subsection{Model B(i) (S-dual model)}
  The cost of S-dual model with (a) model is obtained  by 
    \begin{eqnarray}
\overrightarrow{C}&=&N(1-p)c +Np x^{(u)}_1 + \sum_{j=1}^4 S_j^{(u)}(N,p,\vec{\delta}) \equiv N(1-p)c + Npx^{(u)}_1 +S^{(u)}(N,p,\vec{\delta}) , \\
\underrightarrow{C}&=&Npc +N(1-p)x^{(d)}_1 -\sum_{j=1}^4 S_j^{(d)}(N,p,\vec{\delta})\equiv  Npc +N(1-p)x^{(d)}_1 - S^{(d)}(N,p,\vec{\delta}) . 
\end{eqnarray}
Moreover we have to  impose the following conditions on (25) and (26) due to a duality condition. 
  \begin{eqnarray}
  \overrightarrow{C} - \underrightarrow{C} &=& Np( x^{(u)}_1-c) +N(1-p)(c-x^{(d)}_1 ) + ( \delta_{1}+ \delta_{1,2}+  \delta_{1,2,3}+ \cdots + \delta_{1,2.\cdots, N-1}   )\nonumber\\
  &=&  Np( x^{(u)}_1-c) +N(1-p)(c-x^{(d)}_1 )+ \sum_{i=1}^{N-1} (N-i)\delta_{(i)}. 
  \end{eqnarray}

  In the case without any bypasses, we can find Nash flow by 
  $ \overrightarrow{C} =\underrightarrow{C} $ where $ S_j^{(u)}(N,p,\vec{\delta}) = S_j^{(d)}(N,p,\vec{\delta})=0 $.
  Then we obtain 
 \begin{eqnarray}
x^{(d)}_1&=&pX -2c p+c\;\; OR\;\;\; p=\frac{x^{(d)}_1 -c}{X-2c},\nonumber \\
  x^{(u)}_1 &=&1-pX +2c p-c.
 \end{eqnarray}
 Moreover we find for $ X=1$
 \begin{equation}
  x^{(u)}_1= \frac{1-p}{p} x^{(d)}_1 +\frac{2p-1}{p} c.  
 \end{equation}
 From the duality condition (27) with $\overrightarrow{C}= \underrightarrow{C} $, we obtain the following Nash-Duality condition;
 \begin{equation}
 \frac{ x^{(u)}_1-c}{     x^{(d)}_1-c} =  \frac{1-p}{p}.   
 \end{equation}
(30) can be also derived only from  (28). 
This means the Nash flow (28) and (29) are consistent with the  duality condition. 
Then the cost $C$ is given by 
   \begin{equation}
  C=\overrightarrow{C}= \underrightarrow{C}= N\Bigr( p(1-p)(1-2c)+c \Bigl).  
 \end{equation}
 When $p=0$ (or  $p=1$), we find $x^{(u)}_1=1-c \;(or\; c)$, $x^{(d)}_1=c \;(or\;1-c)$ and $C=cN$. 
  When $p=1/2$, we get $ x^{(u)}_1= x^{(d)}_1=X/2$ and  $ C= \frac{N}{4}(1+2c)$.  
 Notice that there are Nash flow only when $0 \leq c \leq X =1$.  
  
In the case with bypasses, a necessary condition for Nash flow is given by
\begin{equation}
N(c+p)-2Ncp-N x^{(d)}_1=S^{(d)}(N,p,\vec{\delta}) -S^{(u)}(N,p,\vec{\delta}). 
\end{equation}  
For special velue of $p$, we obtain
\begin{eqnarray}
Nc-N x^{(d)}_1&=&S^{(d)}(N,p,\vec{\delta}) -S^{(u)}(N,p,\vec{\delta}),   \mbox{ for } p=0\;or1 \nonumber \\ 
\frac{1}{2} N-N x^{(d)}_1&=&S^{(d)}(N,p,\vec{\delta}) -S^{(u)}(N,p,\vec{\delta}),   \mbox{ for } p=\frac{1}{2}.
\end{eqnarray}  
Thus the condition that the paradox occurs are 
\begin{eqnarray}
-Np(1-p) (1-2c)&>&S^{(d)}(N,p,\vec{\delta})=0,   \mbox{ for } p=1 \nonumber \\ 
Np(1-p) (1-2c)&<&S^{(u)}(N,p,\vec{\delta})=0,   \mbox{ for } p=0 \nonumber \\ 
\frac{1}{2} N(\frac{1}{2}-c)(2p-1)^2&>&0 ,   \mbox{ for } p=\frac{1}{2}. 
\end{eqnarray}
that suggest that the paradox does not occur when relevant values of $p$ are substituted into (34).    


\subsection{Strong dual Model}
In this case, every cost function of  upper edges is revarse to the one of the corresponding lower edges, that is to say $(flow\; x)\Leftrightarrow (constant\;c)$. 
A dual flow is defined as the path where cost functions on the upper edges are $flow \;x$, the real flow is $c$ and also 
where cost functions on the upper edges are $const =c$, the real is $1-c$. 
The costs and the flows of the lower paths are their opposite. 

We can assume that Nash flow in this model is the  dual flow in the case without bypasses or $r=0$ in (a) model.     
This can be proven through  the example in the minimal network in the subsection 2.1.
Then the flow on the edges with cost  function$=c$ is $1-c$ and the one on the edges with the cost function$=x$ is $c$, and 
the total cost from the start node to the terminal one becomes $Nc$ in each path without cost on bypasses. 

Then we obtain 
\begin{equation}
\Delta C= Nc -N\Bigl( p(1-p)(1-2c)+c \Bigr) =p(1-p)(2c-1).
\end{equation}
So we see that Braess' paradox can occur at $c>\frac{1}{2}$ and $p\neq 0,1$. 
This will be also established by numerical analyses in the next section. 
Nash flows for the all minimal networks with 4-nodes and $c=1$ are given in Appendix B. 
 
 \subsection{B(ii) model (T-dual model) Model(ii)} 
 The cost of Model B(i) is obtained  by making the following replacement for (25) and (26);
   \begin{eqnarray}
 \ell &\Longrightarrow& \ell -1, \nonumber \\
  m &\Longrightarrow& m -1, \nonumber \\
N &\Longrightarrow& N-2 \label{eq:36}
  \end{eqnarray}
 But $\ell$ is unchanged in $\delta^{(\ell)}$.  
Moreover 
\begin{equation}
p_{(ii)}=\frac{Np_{(i)}-1}{N-2}, \label{eq:37}
\end{equation}
where $p_{(i)}$ and  $p_{(ii)}$ are the probabilities, corresponding to $p$ in the previous subsection, for the model (i) and the model (ii), respectively. 
Thus the following respacement for the pobabilities are needed in the costs for the upper edge anf the lower edge;
\begin{eqnarray}
Np_{(i)} &=& (N-2)p_{(ii)}+1,\\
N(1- p_{(i)})&=&(N-2)(1-p_{(ii)} )+1
\end{eqnarray}

 So the costs of both paths are  
  \begin{eqnarray}
\overrightarrow{C}=(N-2)(1-p_{(ii)})c +(N-2)p_{(ii)} x^{(u)}_1 +c+  x^{(u)}_1 +S^{(u)\prime}(N-2,p,\vec{\delta}) , \\
\underrightarrow{C}=(N-2)p_{(ii)}c +(N-2)N(1-p_{(ii)})x^{(d)}_1+  c+  x^{(d)}_1 -S^{(d)\prime}(N-2,p,\vec{\delta}),
\end{eqnarray}
where $ S^{(u)\prime}(N-2,p,\vec{\delta})$ and $S^{(d)\prime} (N-2,p,\vec{\delta})$ are constucted by
\begin{equation}
 S^{(u,d)}_{j(\ell-1)} (N-2,p,\vec{\delta})
\end{equation}
instead of $ S^{(u,d)}_{j(\ell)} (N,p,\vec{\delta})$. 

When there are no bypasses, Nash flow is given by 
  \begin{eqnarray}
x^{(u)}_1 =\frac{(2p_{(ii)}-1)(2c-1)}{N} +c+p-2pc,\\
x^{(d)}_1 =1-\frac{(2p_{(ii)}-1)(2c-1)}{N} -c-p+2pc. 
\end{eqnarray}
The cost is given by 
\begin{equation}
C=\overrightarrow{C}= \underrightarrow{C}=Nc +(1-2c)\Bigl(  Np(1-p)+ (2p-1)^2 \frac{N+1}{N} \Bigr).
\end{equation}
The cost with bypasses is $Nc$ under the same hypothesis in the previous subsection. 
So the difference between the costs with bypasses and without bypass is given by 
\begin{equation}
\Delta C=-(1-2c)\Bigl(  Np(1-p)+ (2p-1)^2 \frac{N+1}{N} \Bigr).
\end{equation}
We also find there may be Braess' paradox under the same condition $c>\frac{1}{2}$ as (35).

    \subsection{A(ii) model (T-symmetric model) }
 The cost of Model A(i) is obtained  by making the following replacement (\ref{eq:36}) and (\ref{eq:37}) for (17) and (18);
  So the costs of both paths are  
  \begin{eqnarray}
\overrightarrow{C}=(N-2)(1-p_{(ii)})c +(N-2)p_{(ii)} x^{(u)}_1 +c+  x^{(u)}_1 +S_j^{(u)\prime}(N-2,p,\vec{\delta}) , \\
\underrightarrow{C}=(N-2)(1-p_{(ii)})c +(N-2)Np_{(ii)}x^{(d)}_1+  c+  x^{(d)}_1 -S_j^{(d)\prime}(N-2,p,\vec{\delta}).
\end{eqnarray}
The condition for Nash flow is 
\begin{equation}
 Np(x^{(u)}_1-x^{(d)}_1)+S_j^{(u)\prime}(N-2,p,\vec{\delta})+S_j^{(d)\prime}(N-2,p,\vec{\delta})=0. \label{eq:49}
\end{equation}
When there are not any bupasses, this equation becomes 
\begin{equation}
 Np(x^{(u)}_1-x^{(d)}_1)=0. 
\end{equation}
So we obtain $ x^{(u)}_1=x^{(d)}_1=\frac{1}{2}$.  
Then the cost is given by 
 \begin{equation}
C= \overrightarrow{C}=\underrightarrow{C} =N\bigl( c(1-p)+\frac{p}{2} \bigr). 
\end{equation}

In the case with bypasses, we obtain from (\ref{eq:49})
\begin{equation}
C= \overrightarrow{C}=\underrightarrow{C} =N \bigl( c(1-p)+px^{(d)}_1 \bigr)+S_j^{(u)\prime}(N-2,p,\vec{\delta}).  
\end{equation}
So we obtain
\begin{equation}
\Delta C=\frac{N p}{2}  (x^{(u)}_1-x^{(d)}_1)+S_j^{(u)\prime}(N-2,p,\vec{\delta}).
\end{equation}
Summarize these, we conjecture the following condition for the paradox  should be fulfilled;
\begin{eqnarray}
S_j^{(u)\prime}(N-2,p,\vec{\delta}) &>&0, \;\;\mbox{ for } p=0, \nonumber \\
S_j^{(u)\prime}(N-2,p,\vec{\delta}) &>&0, \;\;\mbox{ for } p=x^{(u)}_1=x^{(d)}_1=\frac{1}{2},  \nonumber \\
\frac{N p}{2}  (x^{(u)}_1-x^{(d)}_1)+S_j^{(u)\prime}(N-2,p,\vec{\delta}) &>&0, \;\;\mbox{ for } p=1.  
\end{eqnarray}

Whether  the paradox occurs or not should be determined by computer simulations. 




\section{Simulations}
We must estimate the costs of $2^{N+1}$ paths to actually find Nadh flow. 
As $N$ becomes larger, it is really impossible to find Nash flows.  
Thus we randomly pick up some paths  and find candidates for Nash flows by imosing the condition that the corresponding costs have the same value.  
In this article, unknown variables are ont only the flows on all edges but also constant cost $c$. 
To solve simultaneous equations as $c$ is an unknown variable is a  distinction in this article. 
The results given by simulations leads some necessary conditions for Nash flow.

Preliminarly we set $N=4$ and $N=5$ to study the difference between odd $N$ and even $N$. 
After gripping the results, we simulate the cases with $N=100$. 
We try simulations of  5$\sim 10$ times, whose trials show the convergence of the results,  for each parameter set. 

\subsection{Positive Nash Flow}
In these simulations, we find that there are essentially no $N$ dependence in results.  
They are summarized in Appendix C. 
There are some cases that can not be identified any Nash flow. 
Such cases also appear in (d) right and (e) right in Table 4 and (i) and (j) in Table 5 of Appendix B. 
In the cases, some paths have to be neglected by the definition of Nash flow\cite{Rou}.  
It is possible to carry out the procedure and identify Nash flows in the simple models in Appendix B. 
The final results are just described in Table 4 and 5.   
It is, however, really impossible to cary out it  in the ladder network because of the complicated model. 
Such situations actually occur in model (b) and (c) as shown in some tables in Appendix C. 

\subsection{Braess' Paradox}
As result of simulation analyses, we can not find any indication of the paradox in model (b) and (c).  
So we show only the results of model (a) where Braess' paradox can definitely occurs.   
All models, however, do not show Braess' paradox at $r=1$. 
The results are summarized in Table 1,2 and 3 based on the detailed results of Appendix C, where the number "1" in the tables means that paradox occurs at $c>\frac{1}{2}$, 
the number "2" in the tables means that paradox does not occur at any $c$ and the number "3" in the tables means 
that 1 or 2 occurs in every trial.  
Notice that the number "1"never does mean that the paradox occurs at any time but paradox can occur only for $c>\frac{1}{2}$. 
These results are basically consistent with theoretical analyses in the section 3.    
The probability reduces to zero as $N\rightarrow \infty$. 
For $N=4,5$, the paradox occurs with relatively large probability. 
We observe that T-dual and T-symmetric models give the same results, and P-symmetric and 
S-dual models also the same ones from these tables. 
The tables show that the paradox is inclined to occur  in model (ii) (T-models) even in the symmetric model and the dual one. 
This fact means that the paradox is inclined to occur when the cost function of first edges both in the lower and the upper is fixed to be arranged  alternately. 
This fact means that the the cost functions of the first edges and the last edges play crucial role. 

We also find that paradox is inclined to occur at $0<p<1$, that is to say, 
the flow $x$ and the constant cost $c$ coexist in the lower edges and the upper ones.  
There are almost not differences in results  in every $N$ and $p$ except for $p=1$ and $p=0$.   
The results at $p=1$ and $p=0$ are the same. 
As discussed in Appendix C.2, "3" in table 2 may reduce to "2" in the end. 
If so, these three tables become same so that parameter $p$-dependences will become disappears.   
But there is the possibility that flows of some paths is zero and then simultaneous equations must be solved apart from these paths. 
Then the new Nash flows may  cause Braess' type paradox. 
 Fig. 7 shows a phase diagrm of parameter areas with respect to Braess' paradox speculated by simulations.

Some general rules that Braess' paradox does not occur are found.  
When the structures such as Fig. 8 appear somewhere in a network, $c=\frac{1}{2}$ and paradox does not occur in all models. 
As $r$ becomes larger, the probability that the structures happen bcomes larger so that the paradox hardly occurs.
Thus the paradox is easier to occur at not large $r$.  
Even more as $N$ becomes larger, structures such as Fig. 8 tend to appear. 
Then the paradox is difficult to occur. 


\begin{table}[htb]
\caption{(a)model with $p=0$}
\label{tb:aa} 
\begin{center}
\begin{tabular}{c|c|c|c}
\hline\hline
\multicolumn{4}{c}{$p$=0} \\
\hline
$r$ & 0 & $0<r<1$ & 1\\
\hline
T-dual & \parbox[m]{30pt}{\strut 1 \strut} & \parbox[m]{30pt}{\strut 1 \strut} & \parbox[m]{30pt}{\strut 2 \strut}\\
\hline
S-dual & \parbox[m]{30pt}{\strut 2 \strut} & \parbox[m]{30pt}{\strut 2 \strut} & \parbox[m]{30pt}{\strut 2 \strut}\\
\hline
T-sym & \parbox[m]{30pt}{\strut 1 \strut} & \parbox[m]{30pt}{\strut 1 \strut} & \parbox[m]{30pt}{\strut 2 \strut}\\
\hline
P-sym & \parbox[m]{30pt}{\strut 2 \strut} & \parbox[m]{30pt}{\strut 2 \strut} & \parbox[m]{30pt}{\strut 2 \strut}\\
\hline0pt
\end{tabular}
\end{center}
\label{table-position}
\end{table}

\begin{table}[!h]
\caption{(a)model with $0<p<1$}
\label{tb:bb} 
\begin{center}
\begin{tabular}{c|c|c|c}
\hline\hline
\multicolumn{4}{c}{$0<p<1$} \\
\hline
$r$ & 0 & $0<r<1$ & 1\\
\hline
T-dual & \parbox[m]{30pt}{\strut 1 \strut} & \parbox[m]{30pt}{\strut 1 \strut} & \parbox[m]{30pt}{\strut 2 \strut}\\
\hline
S-dual & \parbox[m]{30pt}{\strut 3 \strut} & \parbox[m]{30pt}{\strut3\strut} & \parbox[m]{30pt}{\strut 2 \strut}\\
\hline
T-sym & \parbox[m]{30pt}{\strut 1 \strut} & \parbox[m]{30pt}{\strut 1 \strut} & \parbox[m]{30pt}{\strut 2 \strut}\\
\hline
P-sym & \parbox[m]{30pt}{\strut 3 \strut} & \parbox[m]{30pt}{\strut 3 \strut} & \parbox[m]{30pt}{\strut 2 \strut}\\
\hline
\end{tabular}
\end{center}
\label{table-position}
\end{table}

\begin{table}[!h]
\caption{(a)model with $p=1$}
\label{tb:cc} 
\begin{center}
\begin{tabular}{c|c|c|c}
\hline\hline
\multicolumn{4}{c}{$p$=1}\\
\hline
$r$ & 0 & $0<r<1$ & 1\\
\hline
T-dual & \parbox[m]{30pt}{\strut 1 \strut} & \parbox[m]{30pt}{\strut 1 \strut} & \parbox[m]{30pt}{\strut 2 \strut}\\
\hline
S-dual & \parbox[m]{30pt}{\strut 2 \strut} & \parbox[m]{30pt}{\strut 2 \strut} & \parbox[m]{30pt}{\strut 2 \strut}\\
\hline
S-sym & \parbox[m]{30pt}{\strut 1 \strut} & \parbox[m]{30pt}{\strut 1 \strut} & \parbox[m]{30pt}{\strut 2 \strut}\\
\hline
P-sym & \parbox[m]{30pt}{\strut 2 \strut} & \parbox[m]{30pt}{\strut 2 \strut} & \parbox[m]{30pt}{\strut 2 \strut}\\
\hline
\end{tabular}
\end{center}
\label{table-position}
\end{table}

\begin{figure}[ht]
\centering
\includegraphics[scale=0.3]{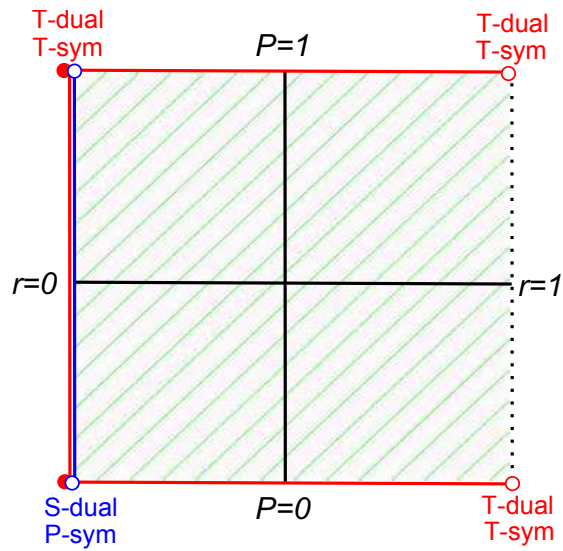}
\caption{Phase diagram for Braess' paradox}
\end{figure}
\begin{figure}[ht]
\centering
\includegraphics[scale=0.3]{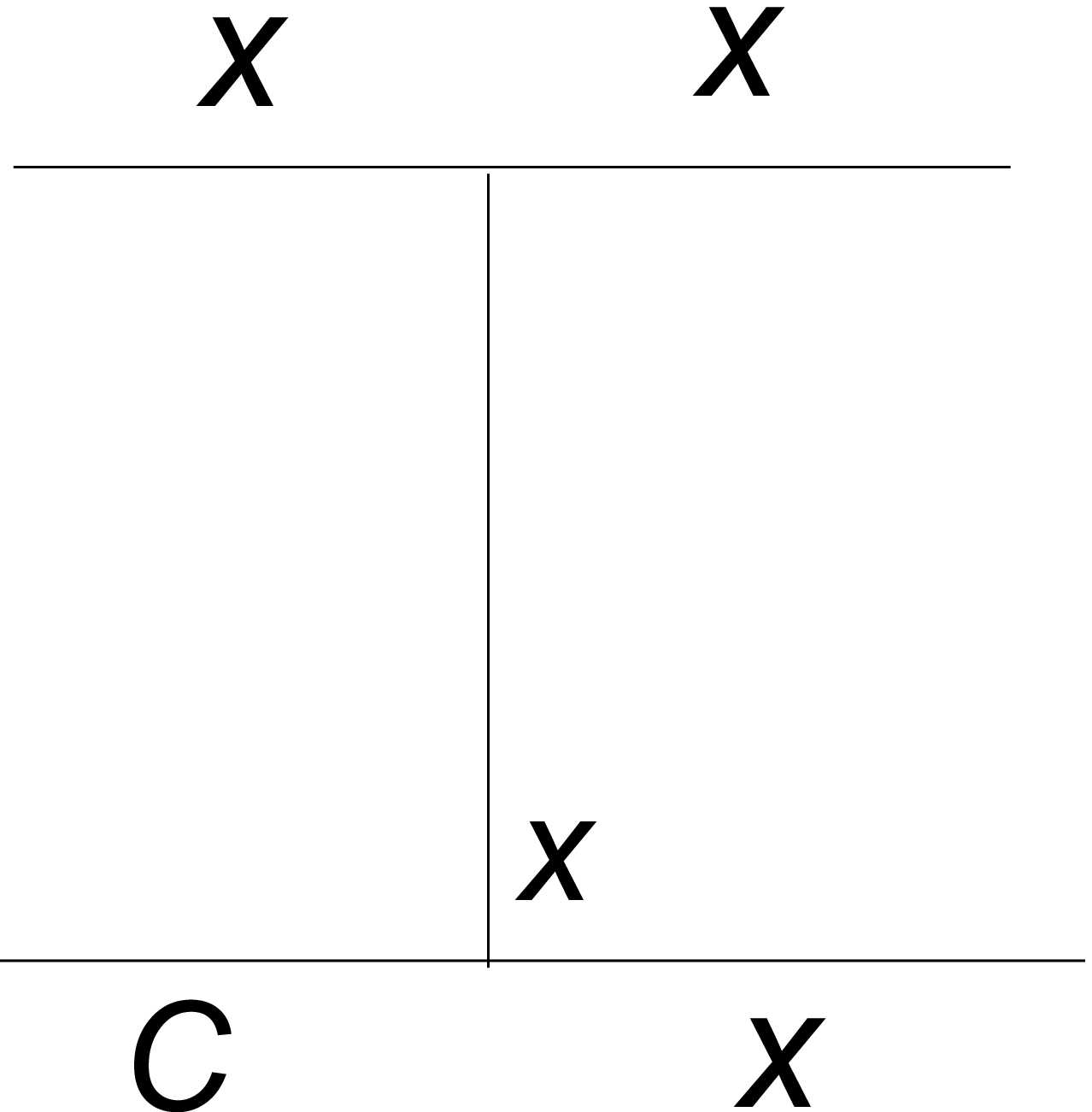}\hspace{10mm}
\includegraphics[scale=0.3]{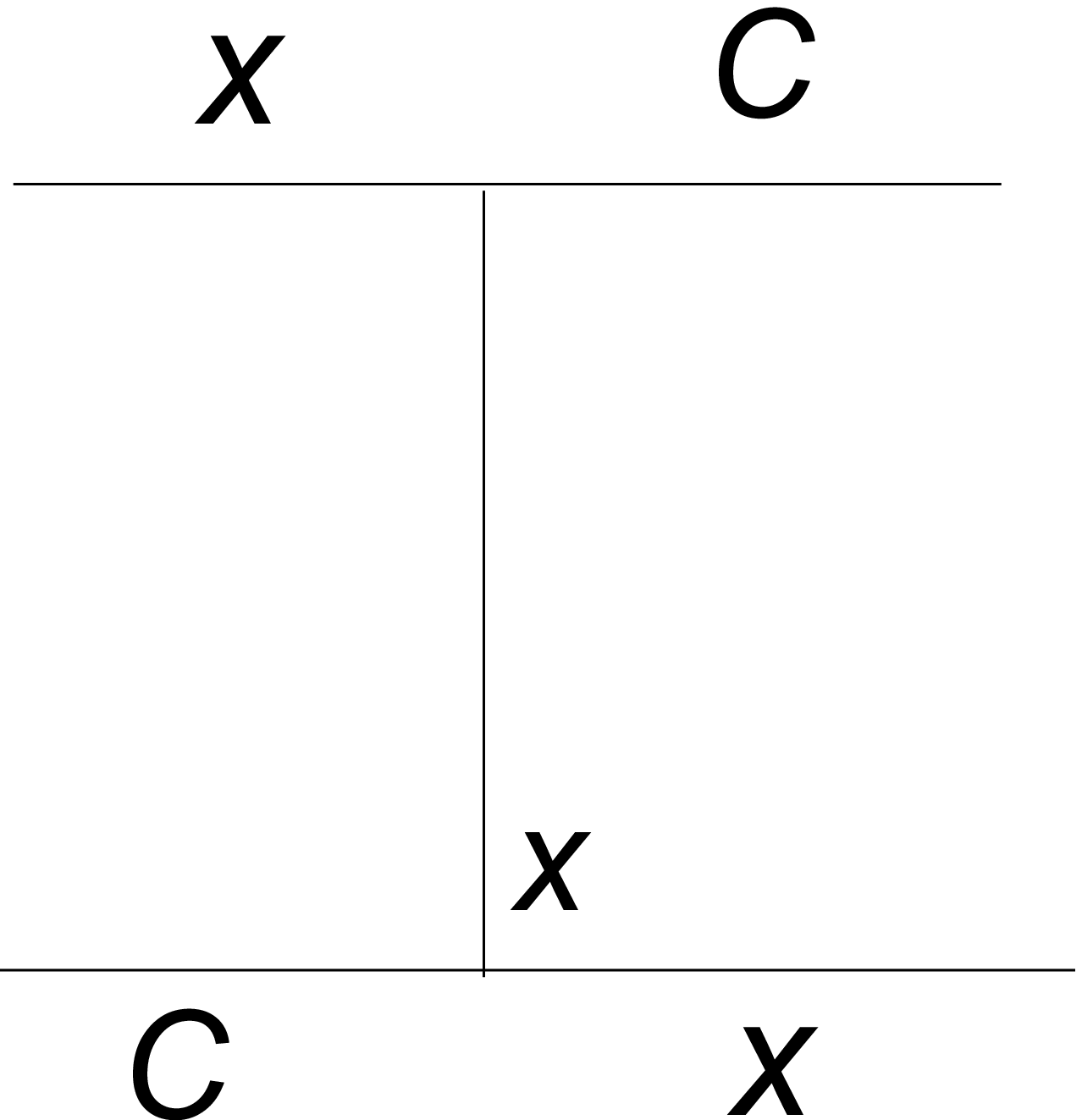}
\caption{General patterns that Braess' paradox does not occur with $c=\frac{1}{2}$. }
\end{figure}

\section{Summaries}

Braess' paradox  \cite{1}, which introducing  new edges (one diagonal line) to networks always does not achieve the efficiency, 
has been originally studied about a traffic flow on a diamond type network.  
Some researchers studied Braess' paradox in the similar type networks by introducing various types of cost functions. 
But whether such paradox occurs or not was not scarcely studied in complex networks except for Dorogovtsev-Mendes network\cite{2}. 
In this article, we study the paradox on Ladder type networks, which is interpreted as Watts-Strogatz network \cite{Watt1}\cite{Watt2} without intersections of bypasses. 
The networks have no longer small world properties, the studies of them can be considered to give the first step to the research about Braess' paradox on Watts and Strogatz type small world network. 
 For the purpose, we construct $4 \times 3$ models as extensions of the original Braess' models in this article.
We theoretically and numerically studied the models on Ladder networks. 
We showed that the both analyses are consistent. 
Last we gave a phase diagram for (a) model, where the cost functions are constant =0 or flow, based on two parameters $r$ and $p$ by numerical simulations. 

\begin{enumerate}
\item This show that paradox hardly occur when the cost of bypasses are itself the flow ($r=1$) on the edge. 
\item The circumstances of the cost functions of the first and the last edges play crusial role in the condition that the paradox occurs. 
\item When the cost functions of the first lower and upper edges are same (constant or the flow itself), 
and the cost functions of the last lower and upper edges are also same, the paradox does not occur. 
\item Further the paradox  is considered  difficult to occur on the whole when the cost functions of bypasses take nonzero constant (model (b) and (c)).
\item The paradox would hardly occur as $N$ becomes larger. 
\end{enumerate}
These facts would give some sugestions for designing effective transportation networks.

\appendix
\setcounter{equation}{0}
 \def\theequation{A$\cdot$\arabic{equation}}
\section{ Proof of Equation (\ref{eq:19})}
We use the following basic formulae in the binomial coefficient due to the proof of (\ref{eq:19})
\begin{eqnarray}
k _nC_k = n \;_{n-1}C_{k-1}, \label{eq:A1}\\
\sum_{r=0}^p \;_nC_r \cdot _mC_{p-r}= _{n+m}C_p, \label{eq:A2}\\
 \; _nC_m=0 \mbox{\;\;for }\;\; n<m \mbox{\;or\;} m<0, \label{eq:A3}
\end{eqnarray}
The last relation is due to the basic formula
\begin{equation}
(-n)! =\frac{n\pi}{n! \sin(n\pi)}.
\end{equation}

 i)$j=1$ case\\
 \begin{eqnarray}
S_1^{(u)}(N,p,\vec{\delta})  _NC_k =\sum_{i=0}^k \delta_{(\ell)} i\;_mC_i\cdot _{\ell}C_{k-i}=\sum_{i=1}^k \delta_{(\ell)} m\;_{m-1}C_{i-1}\cdot _{\ell}C_{k-i} 
\;\mbox{ from  (\ref{eq:A1}) } \nonumber \\ 
 =\sum_{i=0}^{k-1} \delta_{(\ell)} m\;_{m-1}C_{i}\cdot _{\ell}C_{k-i-1} =  \delta_{(\ell)} m\;_{\ell+ m-1}C_{k-1} = \delta_{(\ell)} m \;_{N-1}C_{k-1} 
\mbox{ from (\ref{eq:A2})}
\end{eqnarray}

 ii)$j=2$ case\\
 \begin{eqnarray}
S_2^{(u)}(N,p,\vec{\delta})  _NC_k &=& \sum_{i=k-\ell}^k \delta_{(\ell)} i\;_mC_i\cdot _{\ell}C_{k-i}=\sum_{i=k-\ell}^{k+(k-\ell)}\delta_{(\ell)} i\;_mC_i\cdot _{\ell}C_{k-i}\;
\mbox{ from  (\ref{eq:A3}) } \nonumber \\ 
&=&\sum_{I=0}^k \delta_{(\ell)} (I+k-\ell)\;_{m}C_{I;k-\ell}\cdot _{\ell}C_{k-I-(k-\ell)} \;\mbox{ from $I=i-(k-\ell)$ }\nonumber \\ 
&=&\sum_{I=0}^\ell  \delta_{(\ell)} (I+k-\ell)\;_{m}C_{I;k-\ell}\cdot _{\ell}C_{\ell-I)} 
\;\mbox{ from  (\ref{eq:A3}) }\nonumber \\ 
&=&\sum_{I=0}^\ell  \delta_{(\ell)} m\;_{m-1}C_{I+k-\ell-1}\cdot _{\ell}C_{\ell-I} = \sum_{I=0}^\ell  \delta_{(\ell)} m\;_{m-1}C_{m-I-k+\ell}\cdot _{\ell}C_{I} \;\mbox{ from  (\ref{eq:A1}) } \nonumber \\ 
&=&  \sum_{I=0}^{N-k} \delta_{(\ell)} m\;_{m-1}C_{N-I-k}\cdot _{\ell}C_{I} \; \;\mbox{ from  (\ref{eq:A3}) } \nonumber \\
& =& \delta_{(\ell)} m\;_{\ell+m-1}C_{N-k}  = \delta_{(\ell)} m\;_{N-1}C_{N-k}= \delta_{(\ell)} m \;_{N-1}C_{k-1} \mbox{ from (\ref{eq:A2})}
\end{eqnarray}%

 ii)$j=3$ case\\
  \begin{eqnarray}
S_3^{(u)}(N,p,\vec{\delta})  _NC_k& &=\sum_{i=0}^m \delta_{(\ell)} i\;_mC_i\cdot _{\ell}C_{k-i}=\sum_{i=0}^m \delta_{(\ell)} m\;_{m-1}C_{i-1}\cdot _{\ell}C_{k-i} 
\;\mbox{ from  (\ref{eq:A1}) } \nonumber \\ 
 &=&\sum_{i=0}^{N-k} \delta_{(\ell)} m\;_{m-1}C_{m-i}\cdot _{\ell}C_{\ell-k-i} =  \delta_{(\ell)} m\;_{\ell+ m-1}C_{\ell+m-k} \;\mbox{ from (\ref{eq:A2})}\nonumber \\
&=& \delta_{(\ell)} m \;_{N-1}C_{N-k}=  \delta_{(\ell)} m \;_{N-1}C_{k-1}
\end{eqnarray}
 
  iv)$j=4$ case\\
 \begin{eqnarray}
S_4^{(u)}(N,p,\vec{\delta})  _NC_k &=& \sum_{i=k-\ell}^m \delta_{(\ell)} i\;_mC_i\cdot _{\ell}C_{k-i}=\sum_{I=0}^{m-(k-\ell)}\delta_{(\ell)} i\;_mC_i\cdot _{\ell}C_{k-i}\;
\mbox{ from  $I=i-(k-\ell)$ } \nonumber \\
&=&\sum_{I=0}^{m-(k-\ell)}\delta_{(\ell)} m\;_{m-1}C_{I+(k-\ell)-1} \cdot_{\ell}C_{\ell-I} 
\;\mbox{ from   (\ref{eq:A1}) } \nonumber \\
&=&\sum_{I=0}^{m-(k-\ell)}  \delta_{(\ell)} m\;_{m-1}C_{m-I-(k-\ell)}\cdot _{\ell}C_{I}  =\delta_{(\ell)} m\;_{\ell+m-1}C_{m-k+\ell}\;  \mbox{ from (\ref{eq:A2})} \nonumber \\
& =&\delta_{(\ell)} m\;_{N-1}C_{N-k}= \delta_{(\ell)} m \;_{N-1}C_{k-1} 
\end{eqnarray}%

For $S^{(d)}$, we need only to the replacement $k\rightarrow k^\prime$ to all expressions. 
 Thus, taking account of $N=k+k^\prime$, we obtain 
\begin{equation}
 \frac{S^{(u)}}{S^{(d)}}=\frac{\;_{N-1}C_{k-1}}{\;_{N-1}C_{k^\prime-1}}  =\frac{\;_{N-1}C_{k-1}}{\;_{N-1}C_{N-k^\prime}}  =\frac{\;_{N-1}C_{k-1}}{\;_{N-1}C_{k}}  =\frac{k}{N-k}=\frac{p}{1-p}.
 \end{equation}
 
  \section{ Braess' Paradox in Dyamond type Graph}
  In AppendixB,  we investigate Braess' paradox and Nash flow when the cost=flow or cost=$c$, which is a constant determined by finding Nash flow, in a bypass, and 
  cost =flow or cost=1 in circumferential edges, respectively. 
 A diagonal line means a bypass in the dyamond type graph. 
  The networks are given by Fig.9 and Fig.10 for two-way flow and one-way flow in the bypass, respectively. 
In Fg.9, red color characters $q$ and $u$ are the flows on the bypass and the one on the first upper edge, respectively.  
The black character $x$, which generally shows the flow on the corresponding edge, and $1$'s show the costs for corresponding edges. 
In Fig.10, the red characters $q$ is the flow that goes from the upper node to the lower node and $r$ is the flow with reverse direction on the bypass. 
Table 4 and 5 show Nash flow and whether Braess' paradox occurs or not for each cost function set on edges. 
The third column shows the solution with positive flows.  
In Table 4 and 5, D and U show to pass an upper path and a lower path, respectively. 
M and W show to pass the lower-bypass-upper pass and uppper-bypass-lower pass. respectively. 
P in the right columm means that Braess' paradox occurs in the case.  
Fig.9 and Table 4 correspond to the cases with two-way bypass.  
 Fig.10 and Table 5 correspond to the cases with one-way bypass. 
There are two Nash flows, which happen due to two-way bypass, for  (d) right and (e) right in Table 4.

\begin{figure}[ht]
\centering
\includegraphics[clip,scale=0.8]{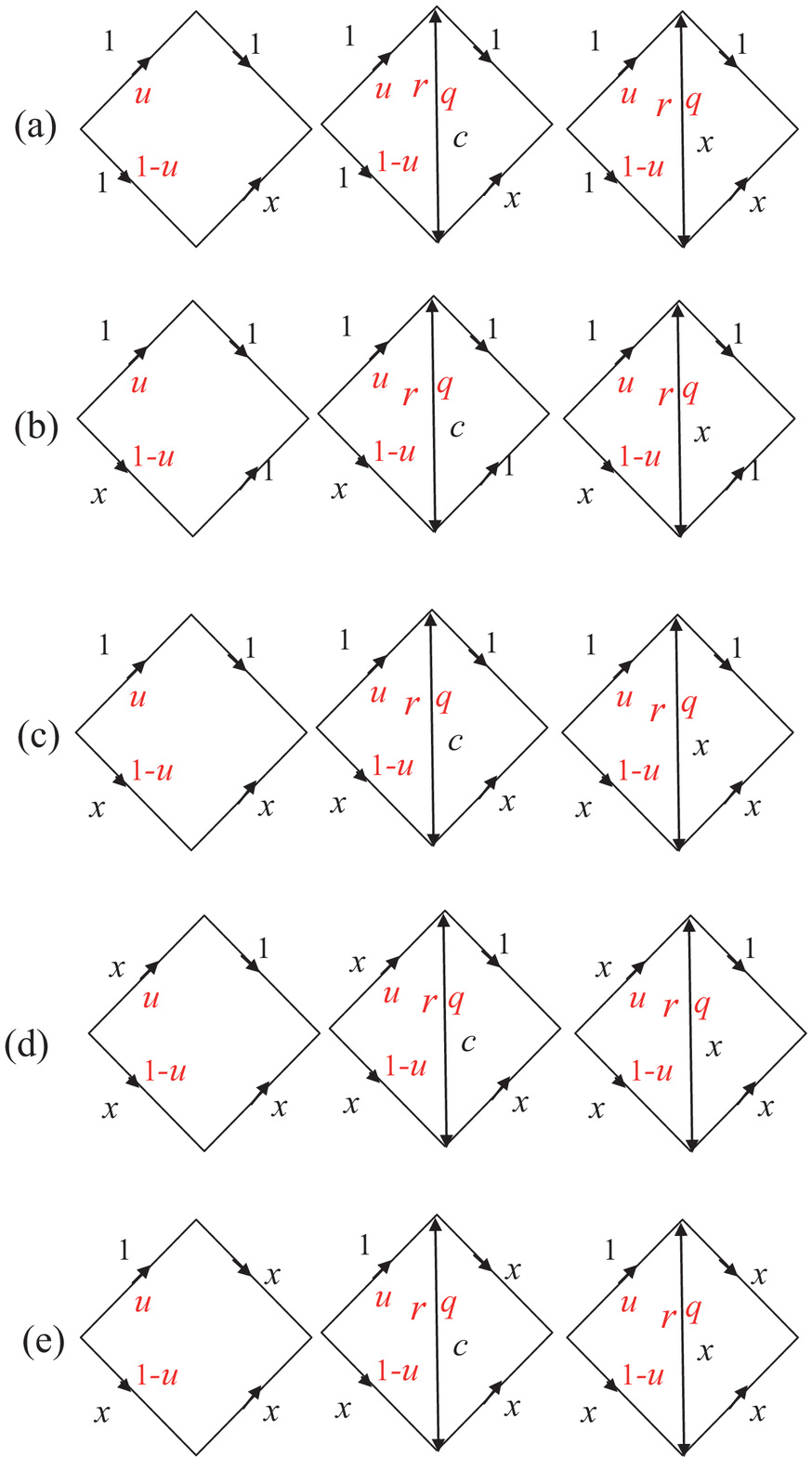}
\caption{Diamond type networks with two-way baypass}
\end{figure}

\begin{table}[t]
\caption{Nash flow and Braess paradox in the cases with one-way bypass}
   \begin{tabular}{|c|c|c|c|c|}\hline
\makebox[5mm]{(a)} &  \makebox[40mm]{Nash Flow}  &  \makebox[35mm]{Positive Nash Flow}  &   \makebox[10mm]{ Flow } & \makebox[15mm]{Paradox} \\ \hline \hline
no bypass & $u=0$ && L & \\ \hline 
midle & $u=q,\; r=c=0$ &same as left  &M or L & none\\ \hline   
 right & $u=q=r=0$  &  &L & none\\ \hline     
 \end{tabular}

   \begin{tabular}{|c|c|c|c|c|}\hline
\makebox[5mm]{(b)} &  \makebox[40mm]{Nash Flow}  &  \makebox[30mm]{Positive Nash Flow}  &     \makebox[10mm]{ Flow }  &\makebox[10mm]{Paradox} \\ \hline \hline
no bypass & $u=0$ && L &\\ \hline 
midle & $u=c=q=0,\; r=arbitrary  $ & same as left   & Dand/orW & none\\ \hline   
 right & $u=q=r=0$  & same as left &L & none\\ \hline     
 \end{tabular}

   \begin{tabular}{|c|c|c|c|c|}\hline
\makebox[5mm]{(c)} &  \makebox[40mm]{Nash Flow}  &  \makebox[35mm]{Positive Nash Flow}  &     \makebox[10mm]{ Flow }  &\makebox[15mm]{Paradox} \\ \hline \hline
no bypass & $u=0$ && L & \\ \hline 
midle & $u=c=q=r=0 $ & same as left   & L & none\\ \hline   
 right & $u=q=r=0$  & same as left &L & none\\ \hline     
 \end{tabular}

   \begin{tabular}{|c|c|c|c|c|}\hline
\makebox[5mm]{(d)} &  \makebox[40mm]{Nash Flow}  &  \makebox[35mm]{Positive Nash Flow}  &     \makebox[10mm]{ Flow }  &\makebox[15mm]{Paradox} \\ \hline \hline
no bypass & $u=\frac{1}{3}$ && L and U &\\ \hline 
midle & $u=q=\frac{1}{2},\;c=r=0 $ & same as left   & L and M & none\\ \hline   
 right & (i)2nd right or (i)1st right & same as left & & none or P\\ \hline     
 \end{tabular}

   \begin{tabular}{|c|c|c|c|c|}\hline
\makebox[5mm]{(e)} &  \makebox[40mm]{Nash Flow}  &  \makebox[30mm]{Positive Nash Flow}  &     \makebox[10mm]{ Flow }  &\makebox[15mm]{Paradox} \\ \hline \hline
no bypass & $u=\frac{1}{3}$ & & U and L & \\ \hline 
midle & $u=c=q=0,\; r=\frac{1}{2} $ & same as left   & L and W & P\\ \hline   
 right & (j)2nd right or (j) 1st right  &same as left  &  & P or none\\ \hline     
 \end{tabular}
 \end{table}

  \begin{table}[b]
\caption{Nash flow and Braess paradox in the cases with two-way bypass}
 \begin{tabular}{|c|c|c|c|c|}\hline
\makebox[5mm]{(f)} &  \makebox[35mm]{Nash Flow}  &  \makebox[40mm]{Positive Nash Flow}  &\makebox[10mm]{ Flow }  &\makebox[15mm]{Paradox} \\ \hline \hline
no bypass & $u=0$ && L & \\ \hline 
1st left & $u=q,\;c=0$ & same as left  & uncertain & none\\ \hline   
 2nd left & $u+q=0,\;c=0$ & same as left &L &none\\ \hline   
 2nd right &$u=q=0$  & same as left & L &none\\ \hline   
1st right & $u=q=0$  &  same as left&L & none\\ \hline     
 \end{tabular}

\begin{tabular}{|c|c|c|c|c|}\hline
\makebox[5mm]{(g)} &  \makebox[35mm]{Nash Flow}  &  \makebox[30mm]{Positive Nash Flow}  &   \makebox[10mm]{ Flow }  &\makebox[15mm]{Paradox} \\ \hline \hline
no bypass & $u=0$ && L & \\ \hline 
1st left & $u=q=c=0$ & same as left  & L & none\\ \hline   
  2nd left & $u=c=0,\;q=uncertain $ &same as left&Dand/orM or N &none\\ \hline   
 2nd right &$u=q=0$  &same as left  & L &none\\ \hline   
1st right & $u=q=0$  & same as left &L & none\\ \hline     
 \end{tabular}
 
 \begin{tabular}{|c|c|c|c|c|}\hline
\makebox[5mm]{(h)} &  \makebox[35mm]{Nash Flow}  &  \makebox[40mm]{Positive Nash Flow}  &  \makebox[10mm]{ Flow } & \makebox[15mm]{Paradox} \\ \hline \hline
no bypass & $u=0$ && L & \\ \hline 
1st left & $u=c,\; q=-2c$ & $c=u=q=0$  &L & none\\ \hline   
 2nd left & $u=-c,\; q=-2c$ & $c=u=q=0$  &L &none\\ \hline   
 2nd right &$u=q=0$  &  & L &none\\ \hline   
1st right & $u=q=0$  &  &L & none\\ \hline     
 \end{tabular}

 \begin{tabular}{|c|c|c|c|c|}\hline
\makebox[5mm]{(i)} &  \makebox[30mm]{Nash Flow}  &  \makebox[40mm]{Positive Nash Flow}  &  \makebox[10mm]{ Flow } & \makebox[10mm]{Paradox} \\ \hline \hline
no bypass & $u=\frac{1}{3}$ && U and B &  \\ \hline 
1st left & $u=\frac{1-c}{2},\; q=\frac{1-3c}{2}$ & same as left   &LandW or N & P for $c<\frac{1}{3}$ \\ \hline   
 2nd left & $u=\frac{c+1}{2},\; q=-\frac{1+3c}{2}$ & same as left  with $1\leq -c\leq \frac{1}{3}$ &B or N &none\\ \hline   
 2nd right & $u=\frac{1}{3},\;q=0$  & same as left & & none\\ \hline 
 1st right &$u=2q=\frac{2}{5}$  &  same as left& Uand/orLand/orM & P\\ \hline     
 \end{tabular}

 \begin{tabular}{|c|c|c|c|c|}\hline
\makebox[5mm]{(j)} &  \makebox[30mm]{Nash Flow}  &  \makebox[40mm]{Positive Nash Flow}  &  \makebox[10mm]{ Flow } & \makebox[15mm]{Paradox} \\ \hline \hline
no bypass & $u=\frac{1}{3} $ && M & \\ \hline 
left & $u=-c,\; q=-\frac{1+3c}{2}$ & same as left with $-1\leq c \leq  \frac{-1}{3} $ & U or  W & none \\ \hline   
2nd left & $u=c,\; q=\frac{1-3c}{2}$ & same as left  with $\frac{1}{3}\geq c\geq \frac{1}{5}$ &U or B or M & P for $ c < \frac{1}{3}$ \\ \hline   
 2nd right & $u=q=\frac{1}{5}$  & same as left& UandLandW& P \\ \hline
1st right &$u=\frac{1}{2} ,q=0$  & same as left &  & none\\ \hline      
 \end{tabular}
\end{table}

\begin{figure}[ht]
\centering
\includegraphics[clip,scale=0.8]{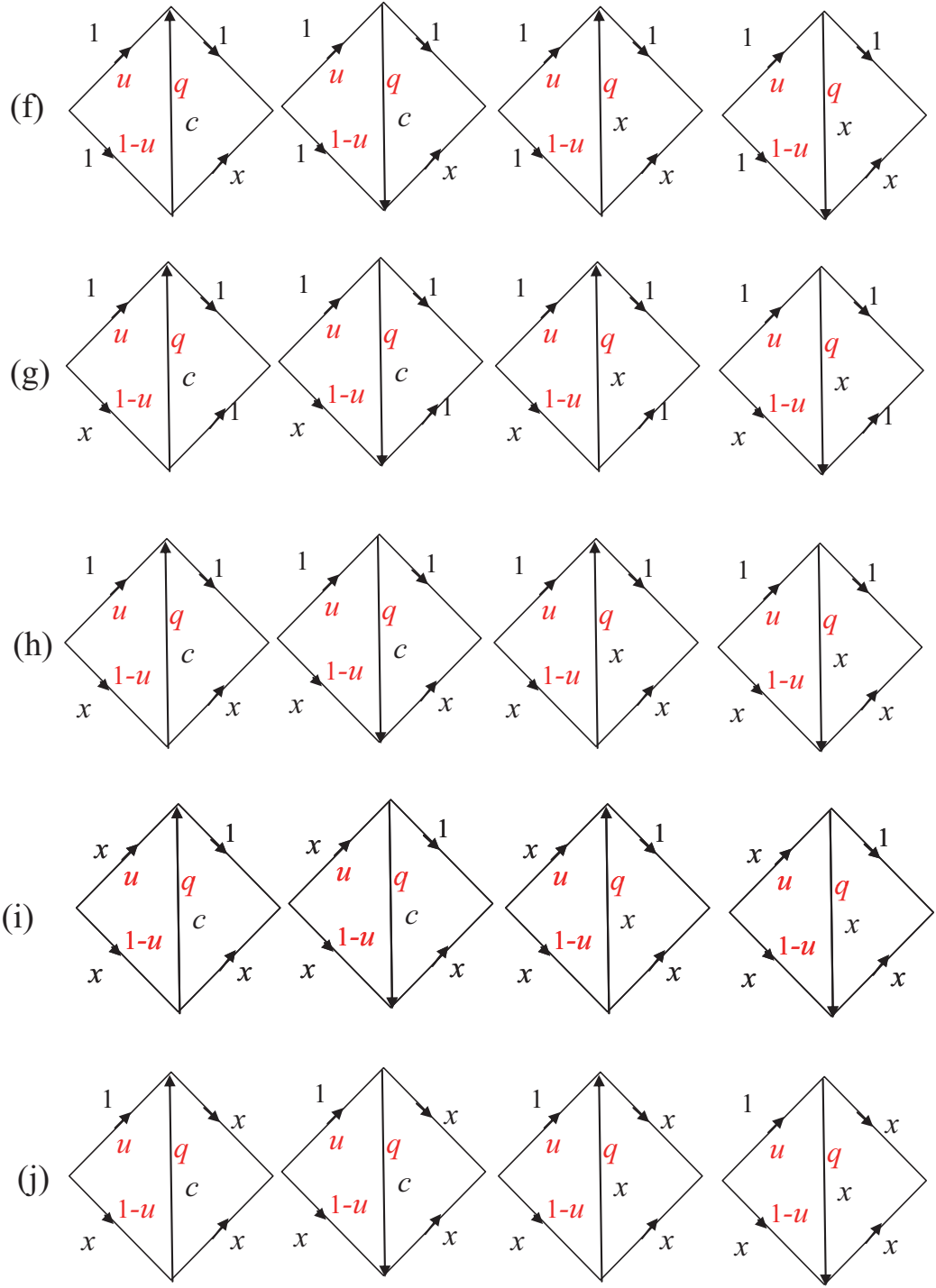}
\caption{Diamond type networks with one-way baypass}
\end{figure}

\clearpage
  \section{ Summary of Nash Flow of all models}
  We show Nash flows of all models in Appendix C. 
  We first give flows on bypasses.  
After that, Nash flows and presence or absence of Braess' paradox are given.  
\subsection{Flow on bypass}  
   Flows on bypasses are divided into the following $3$ types;
 
1. $\delta^\pm$ is arbitrary in all bypasses. 


2. $\delta=\delta^+=\delta^-=0 $  in all bypasses. 


3. Mixed type where the flow depends on the costs of two bypasses and the ones of its nearby  four upper and lower edges of a network.  
They are divided into $64(=2^3 \times 2^3)$ possible patterns given by the combinations of $8$ figures as given below. 
The details of them are given as follows where the symbol $xy,(b,c,d;e,f,g)\Rightarrow ij$ shows that $b,c,d$, which takes $x$(flow) or $c$(constant cost), 
are the costs of an upper edge, bypass, a lower edge in a first figure of the combination, respectively. 
$x$ is the corresponding figure name A,B, $\cdots$ H. 
$d,e,f$ and $y$ are the costs of the subsequent edges and the correponding figure name, respectively.  
The numbers $i$ and $j$ taking from 0 to 4 mean the following bypass flows;\\

\noindent
\underline{The meanings of $i,j$}\\
$j=0$ means the bypass flow is arbitrary\\
$j=1$ means that the flow of a bypass is adjusted to the flows as the upper edge and the lower edge are the dual flow. \\
$j=2$ means that  the flows of  a bypass  is adjusted to  the flows as the upper edge and the lower edge take the same value $\frac{1}{2}$.\\
$j=3$ means that the flow of a bypass is $0$, that is $\delta=\delta^+=\delta^-=0 $.  \\
$j=4$ means that the flow of a bypass is $\delta^+=\delta^-\neq 0 $ and $\delta=0$.   \\

\noindent
AA,$(c,c,c;c,c,c)$$\Rightarrow$$00$ \hspace{15pt} AB,$(c,c,c;c,c,x)$$\Rightarrow$$01$ \hspace{15pt} AC,$(c,c,c;x,c,c)$$\Rightarrow$$01$ \hspace{15pt} AD,$(c,c,c;x,c,x)$$\Rightarrow$$02$\\
AE,$(c,c,c;c,x,c)$$\Rightarrow$$03$ \hspace{15pt} AF,$(c,c,c;c,x,x)$$\Rightarrow$$13$ \hspace{15pt} AG,$(c,c,c;x,x,c)$$\Rightarrow$$13$ \hspace{15pt} AH,$(c,c,c;x,x,x)$$\Rightarrow$$23$\\

\noindent
BA,$(c,c,x;c,c,c)$$\Rightarrow$$10$ \hspace{15pt} BB,$(c,c,x;c,c,x)$$\Rightarrow$$11$ \hspace{15pt} BC,$(c,c,x;x,c,c)$$\Rightarrow$$11$ \hspace{15pt} BD,$(c,c,x;x,c,x)$$\Rightarrow$$12$\\
BE,$(c,c,x;c,x,c)$$\Rightarrow$$13$ \hspace{15pt} BF,$(c,c,x;c,x,x)$$\Rightarrow$$13$ \hspace{15pt} BG,$(c,c,x;x,x,c)$$\Rightarrow$$23$ \hspace{15pt} BH,$(c,c,x;x,x,x)$$\Rightarrow$$23$\\

\noindent
CA,$(x,c,c;c,c,c)$$\Rightarrow$$10$ \hspace{15pt} CB,$(x,c,c;c,c,x)$$\Rightarrow$$11$ \hspace{15pt} CC,$(x,c,c;x,c,c)$$\Rightarrow$$11$ \hspace{15pt} CD,$(x,c,c;x,c,x)$$\Rightarrow$$12$\\
CE,$(x,c,c;c,x,c)$$\Rightarrow$$13$ \hspace{15pt} CF,$(x,c,c;c,x,x)$$\Rightarrow$$23$ \hspace{15pt} CG,$(x,c,c;x,x,c)$$\Rightarrow$$13$ \hspace{15pt} CH,$(x,c,c;x,x,x)$$\Rightarrow$$23$\\

\noindent
DA,$(x,c,x;c,c,c)$$\Rightarrow$$20$ \hspace{15pt} DB,$(x,c,x;c,c,x)$$\Rightarrow$$21$ \hspace{15pt} DC,$(x,c,x;x,c,c)$$\Rightarrow$$21$ \hspace{15pt} DD,$(x,c,x;x,c,x)$$\Rightarrow$$24$\\
DE,$(x,c,x;c,x,c)$$\Rightarrow$$23$ \hspace{15pt} DF,$(x,c,x;c,x,x)$$\Rightarrow$$23$ \hspace{15pt} DG,$(x,c,x;x,x,c)$$\Rightarrow$$23$ \hspace{15pt} DH,$(x,c,x;x,x,x)$$\Rightarrow$$23$\\

\noindent
EA,$(c,x,c;c,c,c)$$\Rightarrow$$30$ \hspace{15pt} EB,$(c,x,c;c,c,x)$$\Rightarrow$$31$ \hspace{15pt} EC,$(c,x,c;x,c,c)$$\Rightarrow$$31$ \hspace{15pt} ED,$(c,x,c;x,c,x)$$\Rightarrow$$32$\\
EE,$(c,x,c;c,x,c)$$\Rightarrow$$23$ \hspace{15pt} EF,$(c,x,c;c,x,x)$$\Rightarrow$$33$ \hspace{15pt} EG,$(c,x,c;x,x,c)$$\Rightarrow$$33$ \hspace{15pt} EH,$(c,x,c;x,x,x)$$\Rightarrow$$33$\\

\noindent
FA,$(c,x,x;c,c,c)$$\Rightarrow$$30$ \hspace{15pt} FB,$(c,x,x;c,c,x)$$\Rightarrow$$31$ \hspace{15pt} FC,$(c,x,x;x,c,c)$$\Rightarrow$$31$ \hspace{15pt} FD,$(c,x,x;x,c,x)$$\Rightarrow$$32$\\
FE,$(c,x,x;c,x,c)$$\Rightarrow$$33$ \hspace{15pt} FF,$(c,x,x;c,x,x)$$\Rightarrow$$23$ \hspace{15pt} FG,$(c,x,x;x,x,c)$$\Rightarrow$$23$ \hspace{15pt} FH,$(c,x,x;x,x,x)$$\Rightarrow$$23$\\

\noindent
GA,$(x,x,c;c,c,c)$$\Rightarrow$$30$ \hspace{15pt} GB,$(x,x,c;c,c,x)$$\Rightarrow$$31$ \hspace{15pt} GC,$(x,x,c;x,c,c)$$\Rightarrow$$31$ \hspace{15pt} GD,$(x,x,c;x,c,x)$$\Rightarrow$$32$\\
GE,$(x,x,c;c,x,c)$$\Rightarrow$$33$ \hspace{15pt} GF,$(x,x,c;c,x,x)$$\Rightarrow$$33$ \hspace{15pt} GG,$(x,x,c;x,x,c)$$\Rightarrow$$33$ \hspace{15pt} GH,$(x,x,c;x,x,x)$$\Rightarrow$$33$\\

\noindent
HA,$(x,x,x;c,c,c)$$\Rightarrow$$30$ \hspace{15pt} HB,$(x,x,x;c,c,x)$$\Rightarrow$$31$ \hspace{15pt} GC,$(x,x,x;x,c,c)$$\Rightarrow$$31$ \hspace{15pt} HH,$(x,x,x;x,c,x)$$\Rightarrow$$34$\\
HE,$(x,x,x;c,x,c)$$\Rightarrow$$33$ \hspace{15pt} HF,$(x,x,x;c,x,x)$$\Rightarrow$$33$ \hspace{15pt} HG,$(x,x,x;x,x,c)$$\Rightarrow$$33$ \hspace{15pt} HH,$(x,x,x;x,x,x)$$\Rightarrow$$33$\\

 \begin{figure}[htb]
\begin{tabular}{cc}
\begin{minipage}{0.25\hsize}
\begin{center}
\fbox{
\includegraphics[width=2.5cm]{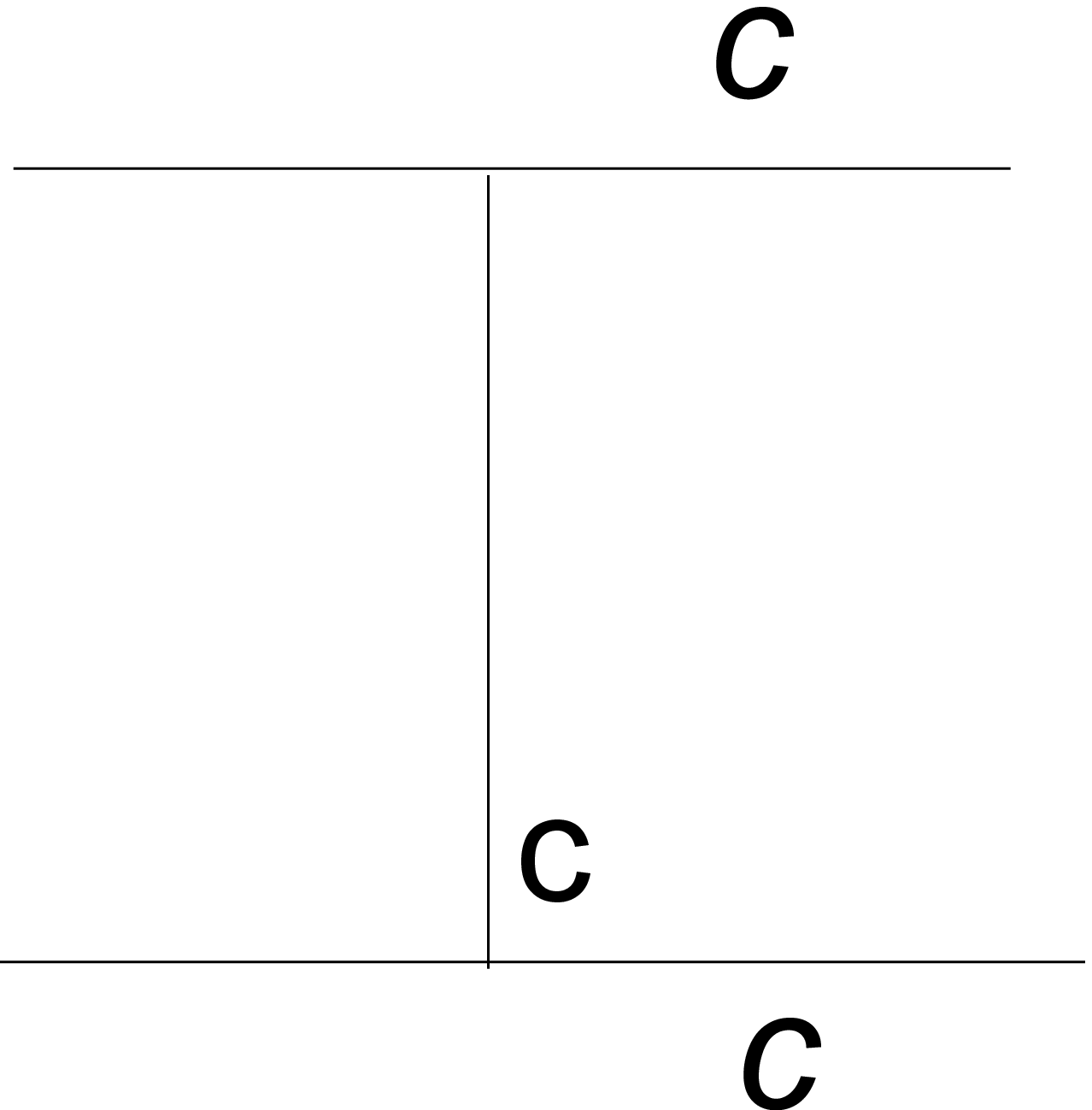}
}
\caption{A}
\label{fig:a}
\end{center}
\end{minipage}
\begin{minipage}{0.25\hsize}
\begin{center}
\fbox{
\includegraphics[width=2.5cm]{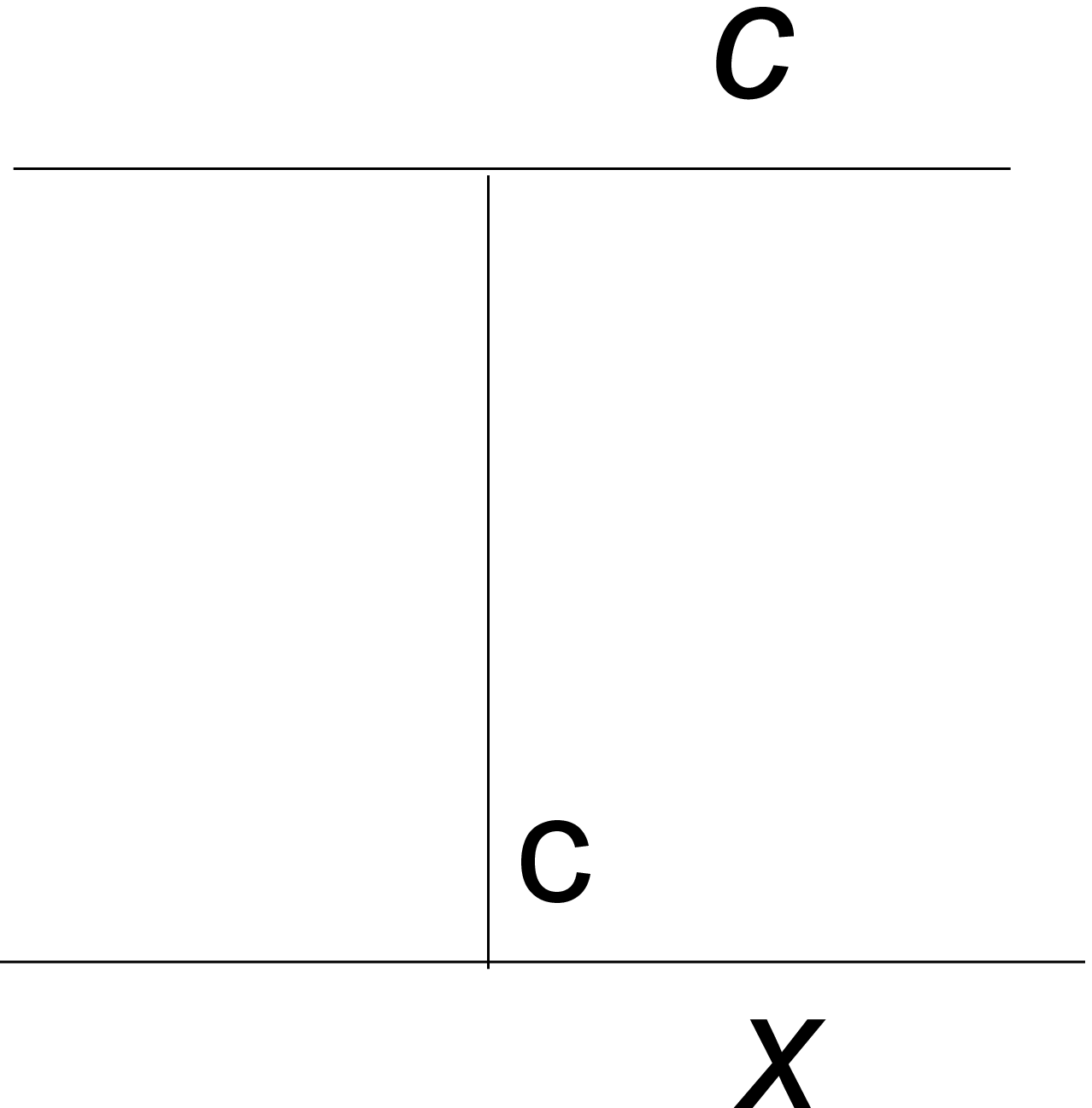}
}
\caption{B}
\label{fig:b}
\end{center}
\end{minipage}
\begin{minipage}{0.25\hsize}
\begin{center}
\fbox{
\includegraphics[width=2.5cm]{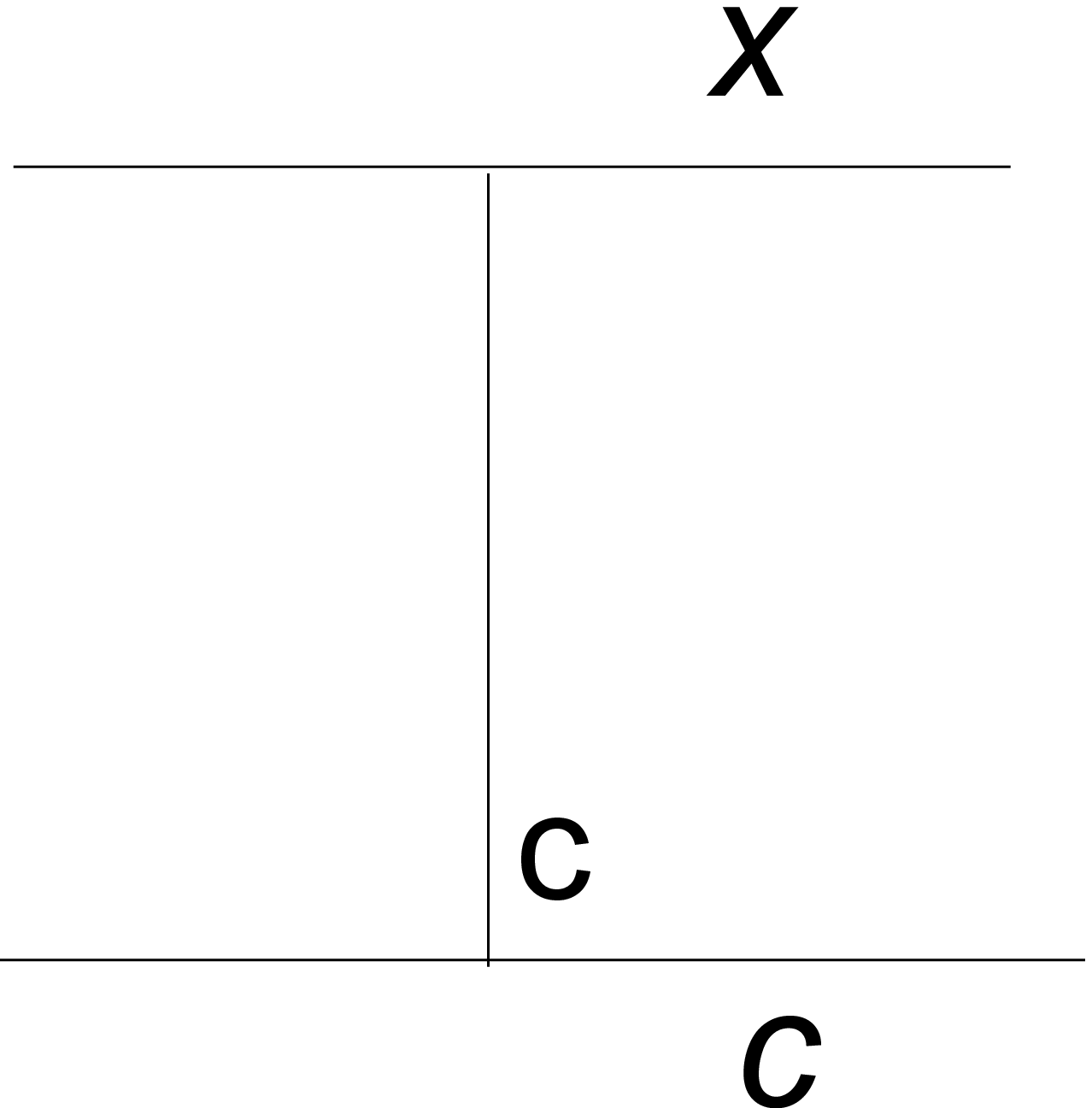}
}
\caption{C}
\label{fig:b}
\end{center}
\end{minipage}
\begin{minipage}{0.25\hsize}
\begin{center}
\fbox{
\includegraphics[width=2.5cm]{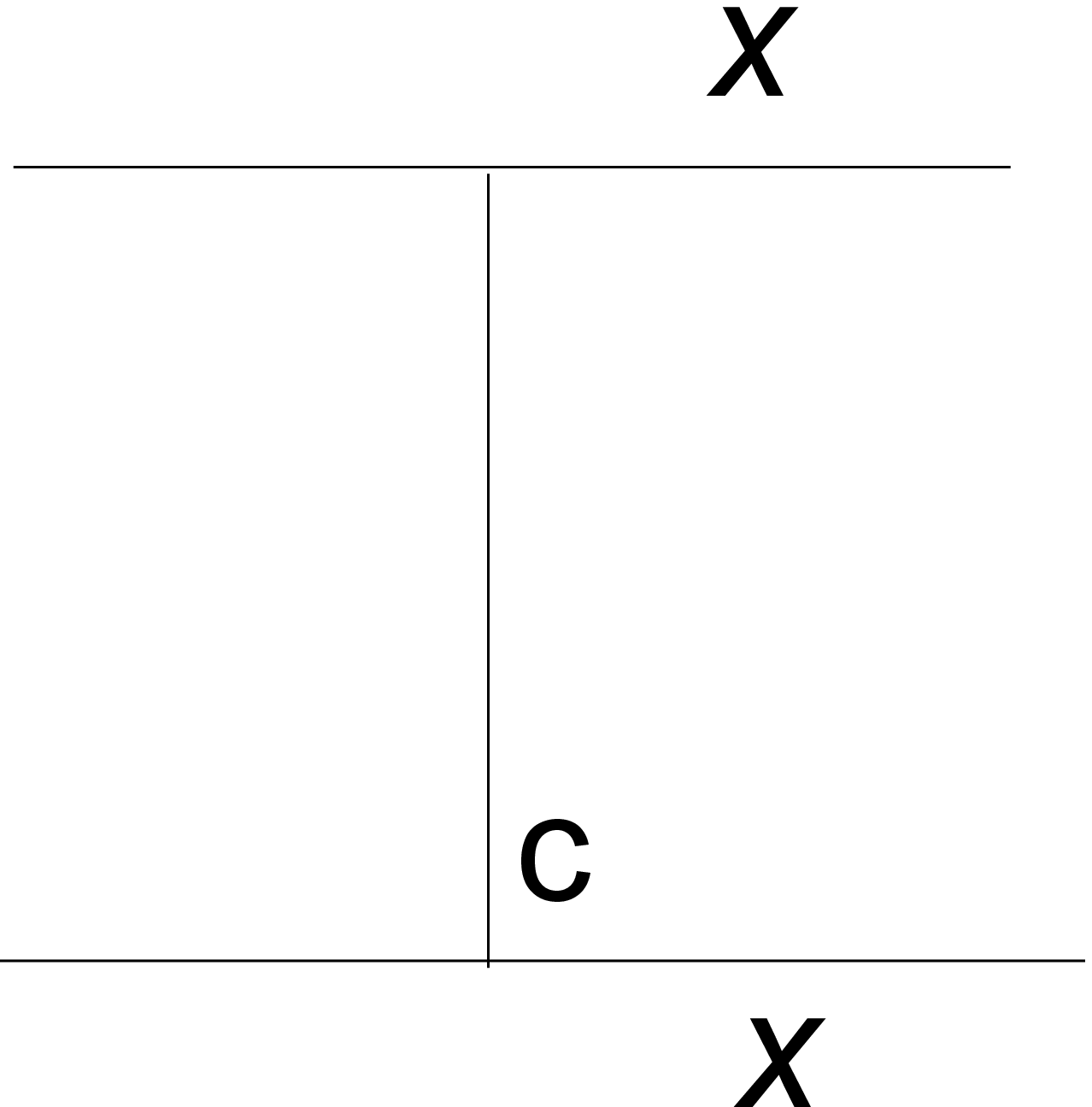}
}
\caption{D}
\label{fig:b}
\end{center}
\end{minipage}
\end{tabular}
\end{figure}

\begin{figure}[htb]
\begin{tabular}{cc}
\begin{minipage}{0.25\hsize}
\begin{center}
\fbox{
\includegraphics[width=2.5cm]{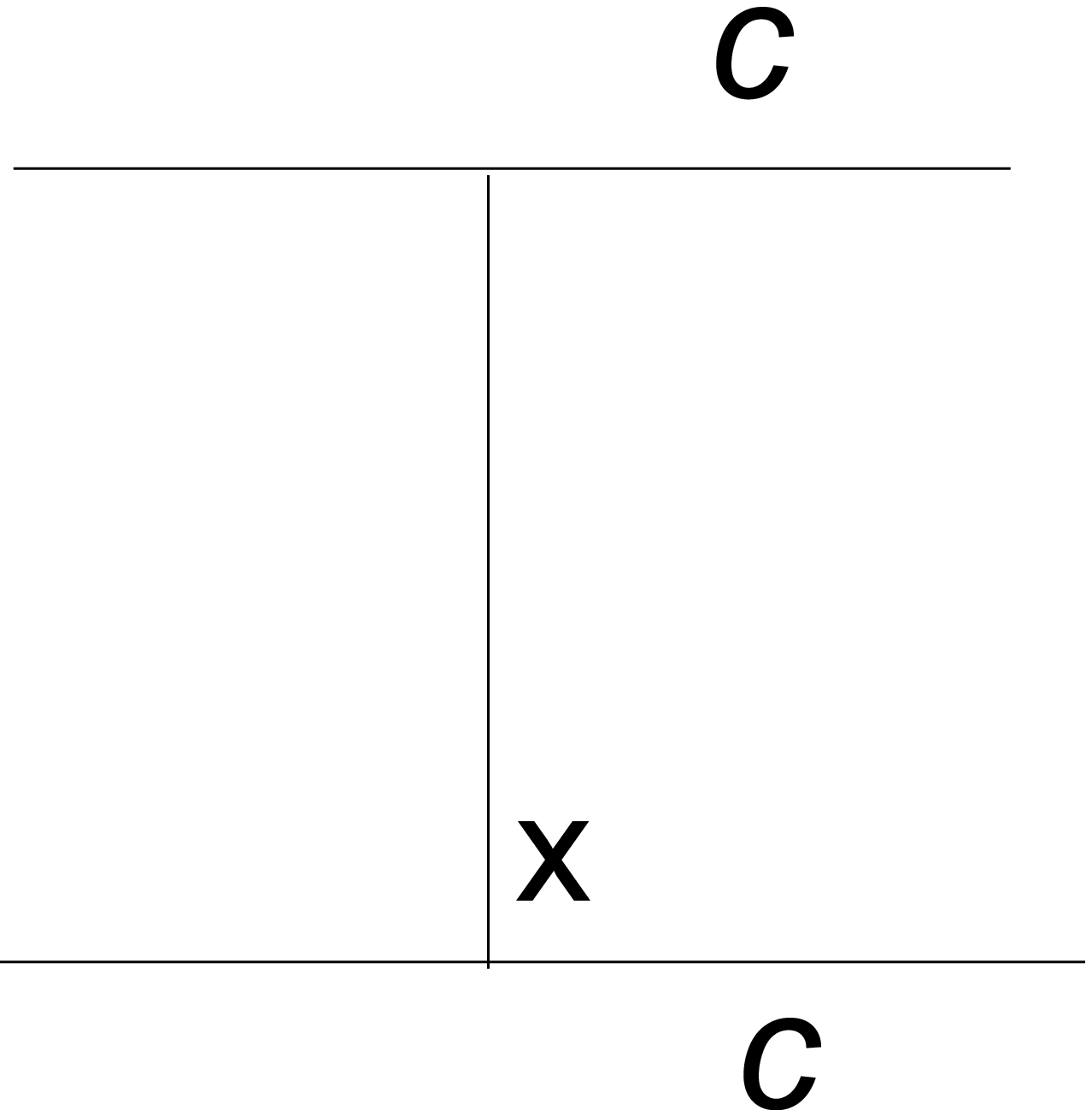}
}
\caption{E}
\label{fig:a}
\end{center}
\end{minipage}
\begin{minipage}{0.25\hsize}
\begin{center}
\fbox{
\includegraphics[width=2.5cm]{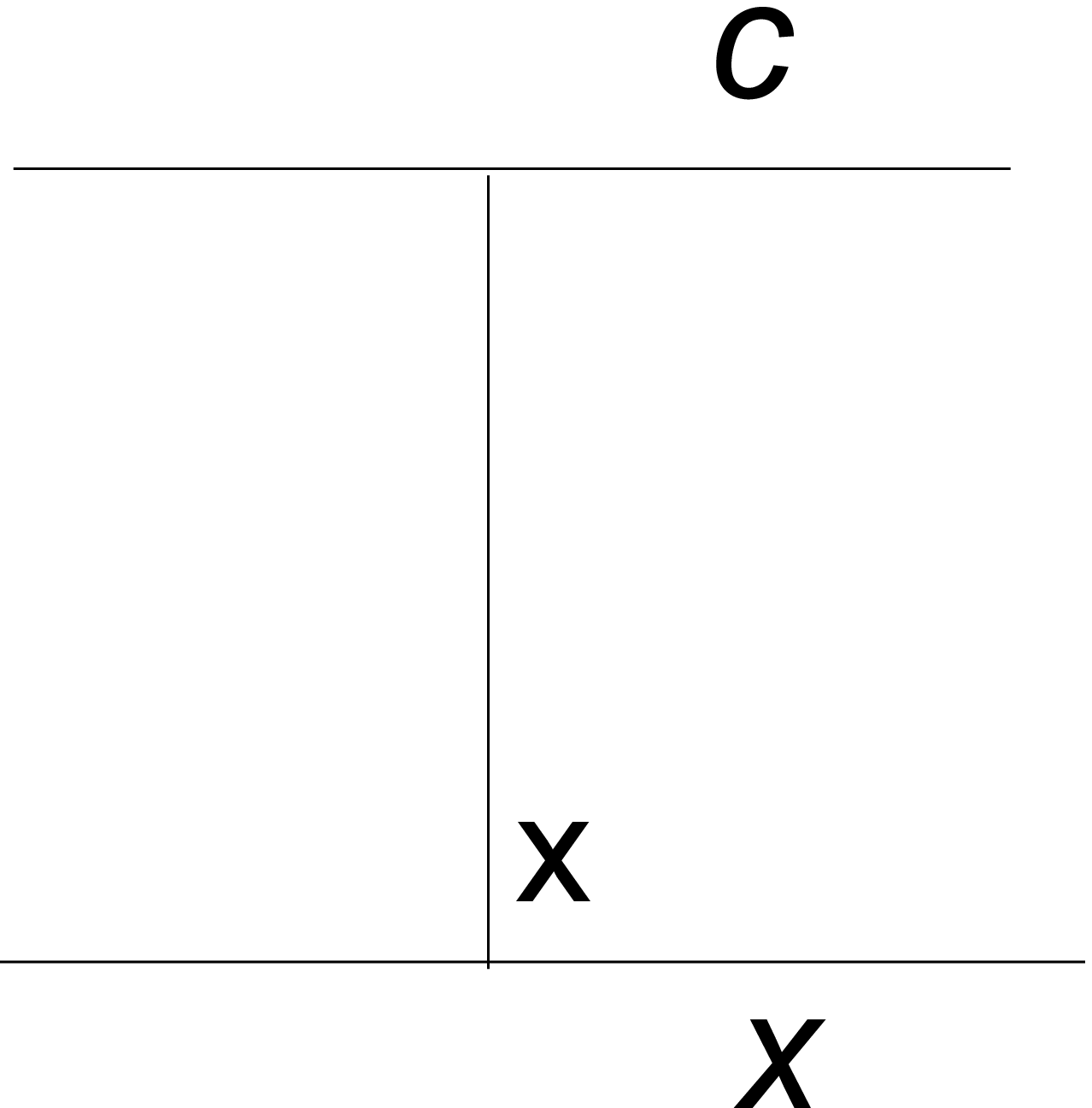}
}
\caption{F}
\label{fig:b}
\end{center}
\end{minipage}
\begin{minipage}{0.25\hsize}
\begin{center}
\fbox{
\includegraphics[width=2.5cm]{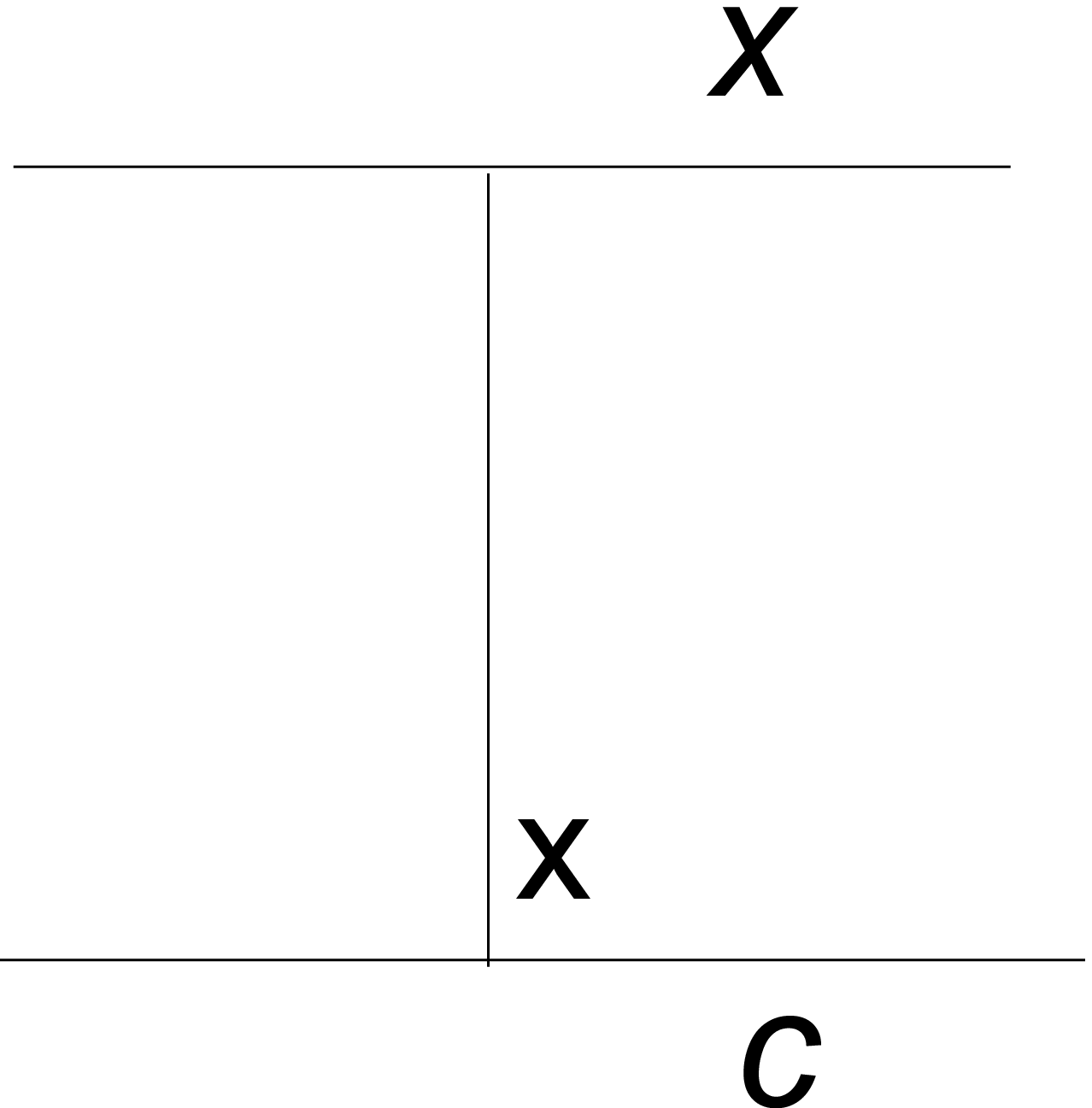}
}
\caption{G}
\label{fig:b}
\end{center}
\end{minipage}
\begin{minipage}{0.25\hsize}
\begin{center}
\fbox{
\includegraphics[width=2.5cm]{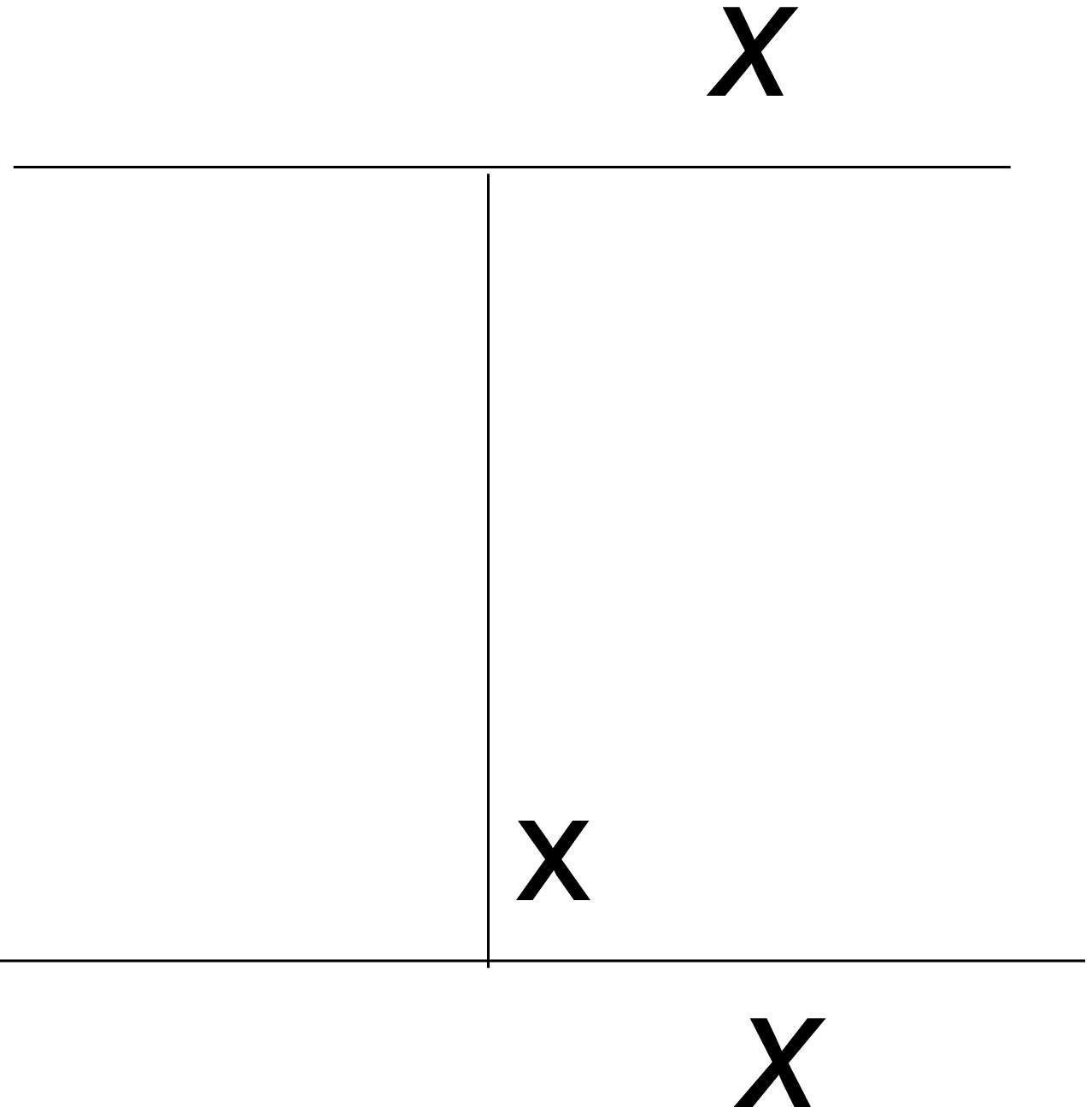}
}
\caption{H}
\label{fig:b}
\end{center}
\end{minipage}
\end{tabular}
\end{figure}

  \subsection{Nash flows and Paradox}  
 We shows Nash flows of all models in the following tables where  "mixed" means that Braess' paradox occurs or not by simulations and  
$\star$  means that one among the following 5 patterns occurs by each simulation. 
  
 1.$c$ and $x_1^{(d)}$ are arbitrary.
 
 2. $x_1^{(d)}=c$ and $c$ is arbitrary;
 
 3.$x_1^{(d)}=1-c$ and $c$ is arbitrary ;
 
 4.$c$ is arbitrary and  $x_1^{(d)}=\frac{1}{2}$;
 
 5.$c=x_1^{(d)}=\frac{1}{2}$;
 
 We speculate that the pattern $\star$ with five patterns reduces to No.5 where $c=x_1^{(d)}=\frac{1}{2}$,   
because we think that 1$\sim$4 happen by taking non-effective paths for finding Nash flows. 
 Then we conjecture that the paradox does not occur in those cases. 
 The columns with plural solutions in these tables can be also speculated to be the same situation as this by the same reason. 
 So we comjecture that Braess' paradox does not occur in these cases. 
 But there is the possibility that flows of some paths is zero and then simultaneous equations must be solved apart from these paths. 
Then the new Nash flows may  cause Braess' type paradox. 
\begin{table}[htbp]
\begin{center}
\begin{tabular}{cc}
\begin{minipage}{0.5\hsize}
\label{tb:1}
\begin{center}
\caption{(a)model with $p$=0 and $r$=0}
\small 
\begin{tabular}{c|c|c|c}
\hline\hline
\multicolumn{4}{c}{(a)model with $p$=0 and $r$=0} \\
\hline
 model  & Nash Flow &  Paradox & $\delta$ \\
\hline
P-sym & \shortstack{$x_1^{(d)}$=arbitrary, \\ $c$=arbitrary} & no & 1 \\
\hline
S-dual & $x_1^{(d)}$=$c$ & no & 3 \\
\hline
T-sym & $x_1^{(d)}$=$1-c$ & yes at $\frac{1}{2}<c$ & 3 \\
\hline
T-dual & $x_1^{(d)}$=$1-c$ & yes at $\frac{1}{2}<c$ & 3 \\
\hline
 \end{tabular}
\label{table-position}
\end{center}
 \end{minipage}
 \begin{minipage}{0.5\hsize}
\label{tb:2}
 \begin{center}
\caption{(a)model with $0<p<1$ and $r$=0}
\small 
\begin{tabular}{c|c|c|c}
\hline\hline
\multicolumn{4}{c}{(a)model with $0<p<1$ and $r$=0} \\
\hline
 model  & Nash Flow &  Paradox & $\delta$ \\
\hline
P-sym & \shortstack{$c$=arbitrary, \\ $x_1^{(d)}$=arbitrary \\ or $c$ or $1-c$ or $\frac{1}{2}$} & mixed & 3 \\
\hline
S-dual & \shortstack{$c$=arbitrary, \\ $x_1^{(d)}$=arbitrary \\ or $c$ or $1-c$ or $\frac{1}{2}$} & mixed & 3 \\
\hline
T-sym & $x_1^{(d)}$=$1-c$ &  yes at $\frac{1}{2}<c$ & 3 \\
\hline
T-dual & $x_1^{(d)}$=$1-c$ &  yes at $\frac{1}{2}<c$ & 3 \\
\hline
 \end{tabular}
\label{table-position}
 \end{center}
 \end{minipage}
 \end{tabular}
 \end{center}
 \end{table} 

\setlength{\baselineskip}{-10pt}

\begin{table}[htbp]
\begin{center}
\begin{tabular}{cc}
\begin{minipage}{0.5\hsize}
\label{tb:3}
\begin{center}
\caption{(a)model with $p$=1 and $r$=0}
\small 
\begin{tabular}{c|c|c|c}
\hline\hline
\multicolumn{4}{c}{(a)model with $p$=1 and $r$=0} \\
\hline
 model  & Nash Flow &  Paradox & $\delta$ \\
\hline
P-sym & \shortstack{$x_1^{(d)}$=$\frac{1}{2}$, \\ $c$=arbitrary} & no & 3 \\
\hline
S-dual & $x_1^{(d)}$=$1-c$ & no & 3 \\
\hline
T-sym & $x_1^{(d)}$=$1-c$ & yes at $\frac{1}{2}<c$ & 3 \\
\hline
T-dual & $x_1^{(d)}$=$1-c$ & yes at $\frac{1}{2}<c$ & 3 \\
\hline
 \end{tabular}
\label{table-position}
\end{center}
 \end{minipage}
 \begin{minipage}{0.5\hsize}
\label{tb:4}
 \begin{center}
\caption{(a)model with $p$=0 and $0<r<1$}
\small 
\begin{tabular}{c|c|c|c}
\hline\hline
\multicolumn{4}{c}{(a)model with $p$=0 and $0<r<1$} \\
\hline
 model  & Nash Flow &  paradox & $\delta$ \\
\hline
P-sym & \shortstack{$c$=arbitrary, \\ $x_1^{(d)}$=arbitrary} & no & 3 \\
\hline
S-dual & $x_1^{(d)}$=$c$ & no & 3 \\
\hline
T-sym & $x_1^{(d)}$=$1-c$ & yes at $\frac{1}{2}<c$ & 3 \\
\hline
T-dual & \shortstack{$x_1^{(d)}$=$1-c$ or \\ $x_1^{(d)}$=$\frac{1}{2}$,$c$=$\frac{1}{2}$} & yes at $\frac{1}{2}<c$ & 3 \\
\hline
 \end{tabular}
\label{table-position}
 \end{center}
 \end{minipage}
 \end{tabular}
 \end{center}
 \end{table}

\begin{table}[htbp]
\begin{center}
\begin{tabular}{cc}
\begin{minipage}{0.5\hsize}
\label{tb:5}
\begin{center}
\caption{(a)model with $0<p<1$ and $0<r<1$}
\small 
\begin{tabular}{c|c|c|c}
\hline\hline
\multicolumn{4}{c}{(a)model with $0<p<1$ and $0<r<1$} \\
\hline
 model  & Nash Flow &  paradox & $\delta$  \\
\hline
P-sym & $\star$ & mixed & 3 \\
\hline
S-dual & $\star$ & mixed & 3 \\
\hline
T-sym & \shortstack{$x_1^{(d)}$=$1-c$ or \\ $x_1^{(d)}$=$\frac{1}{2}$,$c$=$\frac{1}{2}$} & yes at $\frac{1}{2}<c$ &  3 \\
\hline
T-dual & \shortstack{$x_1^{(d)}$=$1-c$ or \\ $x_1^{(d)}$=$\frac{1}{2}$,$c$=$\frac{1}{2}$} & yes at $\frac{1}{2}<c$ &  3 \\
\hline
 \end{tabular}
\label{table-position}
\end{center}
 \end{minipage}
 \begin{minipage}{0.5\hsize}
\label{tb:6}
 \begin{center}
\caption{(a)model with $p$=1 and $0<r<1$}
\small 
\begin{tabular}{c|c|c|c}
\hline\hline
\multicolumn{4}{c}{(a)model with $p$=1 and $0<r<1$} \\
\hline
 model  & Nash Flow &  paradox & $\delta$ \\
\hline
P-sym & \shortstack{$c$=arbitrary, \\ $x_1^{(d)}$=$\frac{1}{2}$} & no & 3 \\
\hline
S-dual & $x_1^{(d)}$=$1-c$ & no & 3 \\
\hline
T-sym & \shortstack{$x_1^{(d)}$=$1-c$ or \\ $x_1^{(d)}$=$\frac{1}{2}$,$c$=$\frac{1}{2}$} & yes at $\frac{1}{2}<c$ &  3 \\
\hline
T-dual & \shortstack{$x_1^{(d)}$=$1-c$ or \\ $x_1^{(d)}$=$\frac{1}{2}$,$c$=$\frac{1}{2}$} & yes at $\frac{1}{2}<c$ &  3 \\
\hline
 \end{tabular}
\label{table-position}
 \end{center}
 \end{minipage}
 \end{tabular}
 \end{center}
 \end{table}

\begin{table}[htbp]
\begin{center}
\begin{tabular}{cc}
\begin{minipage}{0.5\hsize}
\label{tb:7}
\begin{center}
\caption{(a)model with $p$=0 and $r$=1}
\small 
\begin{tabular}{c|c|c|c}
\hline\hline
\multicolumn{4}{c}{(a)model with $p$=0 and $r$=1}\\
\hline
 model  & Nash Flow &  paradox & $\delta$ \\
\hline
P-sym & \shortstack{$c$=arbitrary, \\ $x_1^{(d)}$=arbitrary} & no & 2 \\
\hline
S-dual & $x_1^{(d)}$=$c$ & no & 2 \\
\hline
T-sym & $x_1^{(d)}$=$\frac{1}{2}$,$c$=$\frac{1}{2}$ & no & 2 \\
\hline
T-dual & $x_1^{(d)}$=$\frac{1}{2}$,$c$=$\frac{1}{2}$ & no & 2 \\
\hline
 \end{tabular}
\label{table-position}
\end{center}
 \end{minipage}
 \begin{minipage}{0.5\hsize}
\label{tb:8}
 \begin{center}
\caption{(a)model with $0<p<1$ and $r$=1}
\small 
\begin{tabular}{c|c|c|c}
\hline\hline
\multicolumn{4}{c}{(a)model with $0<p<1$ and $r$=1}\\
\hline
 model  & Nash Flow &  paradox & $\delta$ \\
\hline
P-sym & $x_1^{(d)}$=$\frac{1}{2}$,$c$=$\frac{1}{2}$ & no & 2 \\
\hline
S-dual & $x_1^{(d)}$=$\frac{1}{2}$,$c$=$\frac{1}{2}$ & no & 2 \\
\hline
T-sym & $x_1^{(d)}$=$\frac{1}{2}$,$c$=$\frac{1}{2}$ & no & 2 \\
\hline
T-dual & $x_1^{(d)}$=$\frac{1}{2}$,$c$=$\frac{1}{2}$ & no & 2 \\
\hline
 \end{tabular}
\label{table-position}
 \end{center}
 \end{minipage}
 \end{tabular}
 \end{center}
 \end{table}

\begin{table}[htbp]
\begin{center}
\begin{tabular}{cc}
\begin{minipage}{0.5\hsize}
\label{tb:9}
\begin{center}
\caption{(a)model with $p$=1 and $r$=1}
\small 
\begin{tabular}{c|c|c|c}
\hline\hline
\multicolumn{4}{c}{(a)model with $p$=1 and $r$=1}\\
\hline
 model  & Nash Flow &  paradox & $\delta$ \\
\hline
P-sym & \shortstack{$c$=arbitrary, \\ $x_1^{(d)}$=$\frac{1}{2}$} & no & 2 \\
\hline
S-dual & $x_1^{(d)}$=$1-c$ & no & 2 \\
\hline
T-sym & $x_1^{(d)}$=$\frac{1}{2}$,$c$=$\frac{1}{2}$ & no & 2 \\
\hline
T-dual & $x_1^{(d)}$=$\frac{1}{2}$,$c$=$\frac{1}{2}$ & no & 2 \\
\hline
 \end{tabular}
\label{table-position}
\end{center}
 \end{minipage}
 \begin{minipage}{0.5\hsize}
\label{tb:10}
 \begin{center}
\caption{(b)model with $0<p<1$ and $0<r<1$}
\small 
\begin{tabular}{c|c|c|c}
\hline\hline
\multicolumn{4}{c}{(b)model with $0<p<1$ and $r<1$}\\
\hline
 model  & Nash Flow &  paradox & $\delta$ \\
\hline
P-sym & - & - & - \\
\hline
S-dual & - & - & - \\
\hline
T-sym & - & - & - \\
\hline
T-dual & - & - & - \\
\hline
 \end{tabular}
\label{table-position}
 \end{center}
 \end{minipage}
 \end{tabular}
 \end{center}
 \end{table}

\begin{table}[htbp]
\begin{center}
\begin{tabular}{cc}
\begin{minipage}{0.5\hsize}
\label{tb:11}
\begin{center}
\caption{(b)model with $p$=0 and $r$=1}
\small 
\begin{tabular}{c|c|c|c}
\hline\hline
\multicolumn{4}{c}{(b)model with $p$=0 and $r$=1}\\
\hline
 model  & Nash Flow &  paradox & $\delta$ \\
\hline
P-sym & \shortstack{$c$=arbitrary, \\ $x_1^{(d)}$=arbitrary} & no & 2 \\
\hline
S-dual & $x_1^{(d)}$=$c$ & no & 2 \\
\hline
T-sym & $x_1^{(d)}$=$\frac{1}{2}$,$c$=$\frac{1}{2}$ & no & 2 \\
\hline
T-dual & $x_1^{(d)}$=$\frac{1}{2}$,$c$=$\frac{1}{2}$ & no & 2 \\
\hline
 \end{tabular}
\label{table-position}
\end{center}
 \end{minipage}
 \begin{minipage}{0.5\hsize}
\label{tb:12}
 \begin{center}
\caption{(b)model with $0<p<1$ and $r$=1}
\small 
\begin{tabular}{c|c|c|c}
\hline\hline
\multicolumn{4}{c}{(b)model with $0<p<1$ and $r$=1}\\
\hline
 model  & Nash Flow &  paradox & $\delta$ \\
\hline
P-sym & $x_1^{(d)}$=$\frac{1}{2}$,$c$=$\frac{1}{2}$ & no & 2 \\
\hline
S-dual & $x_1^{(d)}$=$\frac{1}{2}$,$c$=$\frac{1}{2}$ & no & 2 \\
\hline
T-sym & $x_1^{(d)}$=$\frac{1}{2}$,$c$=$\frac{1}{2}$ & no & 2 \\
\hline
T-dual & $x_1^{(d)}$=$\frac{1}{2}$,$c$=$\frac{1}{2}$ & no & 2 \\
\hline
 \end{tabular}
\label{table-position}
 \end{center}
 \end{minipage}
 \end{tabular}
 \end{center}
 \end{table}

\clearpage

\begin{table}[htbp]
\begin{center}
\begin{tabular}{cc}
\begin{minipage}{0.5\hsize}
\label{tb:13}
\begin{center}
\caption{(b)model with $p$=1 and $r$=1}
\small 
\begin{tabular}{c|c|c|c}
\hline\hline
\multicolumn{4}{c}{(b)model with $p$=1 and $r$=1}\\
\hline
 model  & Nash Flow &  paradox & $\delta$ \\
\hline
P-sym & \shortstack{$c$=arbitrary, \\ $x_1^{(d)}$=$\frac{1}{2}$} & no & 2 \\
\hline
S-dual & $x_1^{(d)}$=$1-c$ & no & 2 \\
\hline
T-sym & $x_1^{(d)}$=$\frac{1}{2}$,$c$=$\frac{1}{2}$ & no & 2 \\
\hline
T-dual & $x_1^{(d)}$=$\frac{1}{2}$,$c$=$\frac{1}{2}$ & no & 2 \\
\hline
 \end{tabular}
\label{table-position}
\end{center}
 \end{minipage}
 \begin{minipage}{0.5\hsize}
\label{tb:14}
 \begin{center}
\caption{(c)model with $p$=0 and $r$=0}
\small 
\begin{tabular}{c|c|c|c}
\hline\hline
\multicolumn{4}{c}{(c)model with $p$=0 and $r$=0}\\
\hline
 model  & Nash Flow &  paradox & $\delta$ \\
\hline
P-sym & \shortstack{$c$=0, \\ $x_1^{(d)}$=arbitrary} & no & 1 \\
\hline
S-dual & $c$=0,$x_1^{(d)}$=0 & no & 3 \\
\hline
T-sym & $c$=0,$x_1^{(d)}$=1 & no & 3 \\
\hline
T-dual & $c$=0,$x_1^{(d)}$=1 & no & 3 \\
\hline
 \end{tabular}
\label{table-position}
 \end{center}
 \end{minipage}
 \end{tabular}
 \end{center}
 \end{table}

\begin{table}[htbp]
\begin{center}
\begin{tabular}{cc}
\begin{minipage}{0.5\hsize}
\label{tb:15}
\begin{center}
\caption{(c)model with $0<p<1$ and $r$=0}
\small 
\begin{tabular}{c|c|c|c}
\hline\hline
\multicolumn{4}{c}{(c)model with $0<p<1$ and $r$=0}\\
\hline
 model  & Nash Flow &  paradox & $\delta$ \\
\hline
P-sym & \shortstack{$c$=0, \\ $x_1^{(d)}$=arbitrary \\ or 0 or 1 or $\frac{1}{2}$} & no & 3 \\
\hline
S-dual & \shortstack{$c$=0, \\ $x_1^{(d)}$=arbitrary \\ or 0 or 1 or $\frac{1}{2}$} & no & 3 \\
\hline
T-sym & $c$=0,$x_1^{(d)}$=1 & no & 3 \\
\hline
T-dual & $c$=0,$x_1^{(d)}$=1 & no & 3 \\
\hline
 \end{tabular}
\label{table-position}
\end{center}
 \end{minipage}
 \begin{minipage}{0.5\hsize}
\label{tb:16}
 \begin{center}
\caption{(c)model with $p$=1 and $r$=0}
\small 
\begin{tabular}{c|c|c|c}
\hline\hline
\multicolumn{4}{c}{(c)model with $p$=1 and $r$=0}\\
\hline
 model  & Nash Flow &  paradox & $\delta$ \\
\hline
P-sym & $c$=0,$x_1^{(d)}$=$\frac{1}{2}$ & no & 3 \\
\hline
S-dual & $c$=0,$x_1^{(d)}$=1 & no & 3 \\
\hline
T-sym & $c$=0,$x_1^{(d)}$=1 & no & 3 \\
\hline
T-dual & $c$=0,$x_1^{(d)}$=1 & no & 3 \\
\hline
 \end{tabular}
\label{table-position}
 \end{center}
 \end{minipage}
 \end{tabular}
 \end{center}
 \end{table}

\begin{table}[htbp]
\begin{center}
\begin{tabular}{cc}
\begin{minipage}{0.5\hsize}
\label{tb:17}
\begin{center}
\caption{(c)model with $p$=0 and $0<r<1$}
\small 
\begin{tabular}{c|c|c|c}
\hline\hline
\multicolumn{4}{c}{(c)model with $p$=0 and $0<r<1$}\\
\hline
 model  & Nash Flow &  paradox & $\delta$ \\
\hline
P-sym & \shortstack{$c$=0, \\ $x_1^{(d)}$=arbitrary} & no & 3 \\
\hline
S-dual & $c$=0,$x_1^{(d)}$=0 & no & 3 \\
\hline
T-sym & $c$=0,$x_1^{(d)}$=1 & no & 3 \\
\hline
T-dual & - & - & - \\
\hline
 \end{tabular}
\label{table-position}
\end{center}
 \end{minipage}
  \begin{minipage}{0.5\hsize}
\label{tb:18}
 \begin{center}
\caption{(c)model with $0<p<1$ and $0<r<1$}
\small 
\begin{tabular}{c|c|c|c}
\hline\hline
\multicolumn{4}{c}{(c)model with $0<p<1$ and $0<r<1$}\\
\hline
 model  & Nash Flow &  paradox & $\delta$ \\
\hline
P-sym & - & - & - \\
\hline
S-dual & - & - & - \\
\hline
T-sym & - & - & - \\
\hline
T-dual & - & - & - \\
\hline
 \end{tabular}
\label{table-position}
 \end{center}
 \end{minipage}
 \end{tabular}
 \end{center}
 \end{table} 

\begin{table}[hbtp]
\begin{center}
\begin{tabular}{cc}
\begin{minipage}{0.5\hsize}
\label{tb:19}
\begin{center}
\caption{(c)model with $p$=1 and $0<r<1$}
\small 
\begin{tabular}{c|c|c|c}
\hline\hline
\multicolumn{4}{c}{(c)model with $p$=1 and $0<r<1$}\\
\hline
 model  & Nash Flow &  paradox & $\delta$ \\
\hline
P-sym & $c$=0,$x_1^{(d)}$=$\frac{1}{2}$ & no & 3 \\
\hline
S-dual & $c$=0,$x_1^{(d)}$=1 & no & 3 \\
\hline
T-sym & - & - & - \\
\hline
T-dual & - & - & - \\
\hline
 \end{tabular}
\label{table-position}
\end{center}
 \end{minipage}
 \begin{minipage}{0.5\hsize}
\label{tb:20}
 \begin{center}
\caption{(c)model with $p$=0 and $r$=1}
\small 
\begin{tabular}{c|c|c|c}
\hline\hline
\multicolumn{4}{c}{(b)model with $p$=0 and $r$=1}\\
\hline
 model  & Nash Flow &  paradox & $\delta$ \\
\hline
P-sym & \shortstack{$c$=arbitrary, \\ $x_1^{(d)}$=arbitrary} & no & 2 \\
\hline
S-dual & $x_1^{(d)}$=$c$ & no & 2 \\
\hline
T-sym & $x_1^{(d)}$=$\frac{1}{2}$,$c$=$\frac{1}{2}$ & no & 2 \\
\hline
T-dual & $x_1^{(d)}$=$\frac{1}{2}$,$c$=$\frac{1}{2}$ & no & 2 \\
\hline
 \end{tabular}
\label{table-position}
 \end{center}
 \end{minipage}
 \end{tabular}
 \end{center}
 \end{table}

\begin{table}[hbtp]
\begin{center}
\begin{tabular}{cc}
\begin{minipage}{0.5\hsize}
\label{tb:21}
\begin{center}
\caption{(c)model with $0<p<1$ and $r$=1}
\small 
\begin{tabular}{c|c|c|c}
\hline\hline
\multicolumn{4}{c}{(c)model with $0<p<1$ and $r$=1}\\
\hline
 model  & Nash Flow &  paradox & $\delta$ \\
\hline
P-sym & $x_1^{(d)}$=$\frac{1}{2}$,$c$=$\frac{1}{2}$ & no & 2 \\
\hline
S-dual & $x_1^{(d)}$=$\frac{1}{2}$,$c$=$\frac{1}{2}$ & no & 2 \\
\hline
T-sym & $x_1^{(d)}$=$\frac{1}{2}$,$c$=$\frac{1}{2}$ & no & 2 \\
\hline
T-dual & $x_1^{(d)}$=$\frac{1}{2}$,$c$=$\frac{1}{2}$ & no & 2 \\
\hline
 \end{tabular}
\label{table-position}
\end{center}
 \end{minipage}
 \begin{minipage}{0.5\hsize}
\label{tb:22}
 \begin{center}
\caption{(b)model with $p$=1 and $r$=1}
\small 
\begin{tabular}{c|c|c|c}
\hline\hline
\multicolumn{4}{c}{(c)model with $p$=1 and $r$=1}\\
\hline
 model  & Nash Flow &  paradox & $\delta$ \\
\hline
P-sym & \shortstack{$c$=arbitrary, \\ $x_1^{(d)}$=$\frac{1}{2}$} & no & 2 \\
\hline
S-dual & $x_1^{(d)}$=$1-c$ & no & 2 \\
\hline
T-sym & $x_1^{(d)}$=$\frac{1}{2}$,$c$=$\frac{1}{2}$ & no & 2 \\
\hline
T-dual & $x_1^{(d)}$=$\frac{1}{2}$,$c$=$\frac{1}{2}$ & no & 2 \\
\hline
 \end{tabular}
\label{table-position}
 \end{center}
 \end{minipage}
 \end{tabular}
 \end{center}
 \end{table}


\newpage

%

\begin{thebibliography}{10}
%
\bibitem{1}
D.Braess, A Nagurney and T. Wakolbinger, "On a Paradox of Traffic Planning", Transporttation Science vol.39, No.4, Nov. (2005) 446-450.
%
\bibitem{pas}
E.I.Pas and S.L. Principio, "Braess' paradox:some new insights", Transportation Res. B31(3) pp.265-276, 1997


\bibitem{Zrer}
V.Zrerovich and E. Avineri, "Braess' Paradox in a General Traffic Network", arXive:1207.3251, 2012


\bibitem{Vali}
G.Valiant and T. Roughtgarden, "Braess' paradox in large random graphs", Proceedings of 7th Annual ACM Conference Electronic Commerce (EC), 296-305, 2006

\bibitem{Bloy}
L.A.K.l.Bloy, "AN INTRODUCTION INTO BRAESS' PARADOX", the Degree of Master of Science at Univ. of South Africa, 2007

\bibitem{2}
S.N. Dorogovtsev and J.F.F. Mendes, "Exactly solvable analogy of small-world networks", arXiv:cond-mat/9907445 Jul.(1999).
%
\bibitem{3} 
D.J.Ashton, T C. Jarret andN.F Johnson, "Effect of congestion costs on shortest paths through complex networks", arXiv:cond-mat/0409059 Nov.(2004).
%
\bibitem{4}
 T C. Jarret, D.J.Ashton, M. Fricker and N.F Johnson, "Interplay between function and structure in complex networks", arXiv:physics/0604183 April (2006).
 
 \bibitem{ToyotaB1}
N.Toyota, gBraess like Paradox on Dorogovtsev-Mendes networkhC ITC-CSCC2013 in Yeosu, Korea, S3-10,June 1-July 3  ,2013
\bibitem{ToyotaB2}N. Toyota, gBraess like Paradox in a Small World NetworkhCPreprint arXiv:1304.4744,2013.
\bibitem{Watt1}D. J. Watts and S. H. Strogatz, "Collective dynamics of 'small-world' networks",@Nature,393, 440-442(1998)
\bibitem{Watt2}D. J. Watts, "Six degree-- The science of a connected age", W.W. Norton and Company, New York (2003)

\bibitem{Rou}T. Roughgarden, "Selfish Routing and the Price of Anarchy", MIT press, Cmbridge, Massachusetts, 2005.
\end{thebibliography}
\end{document}